\documentclass[aps,prc,superscriptaddress,showpacs,nofootinbib,floatfix,twocolumn]{revtex4}

\usepackage{graphicx}   
\usepackage{setspace}
\usepackage{rotating}
\usepackage{dcolumn}
\usepackage{bm}
\usepackage{epsfig}
\usepackage{amssymb}
\usepackage{multirow}
\usepackage{CJK}

\begin{document}
\begin{CJK*}{GB}{}
\title{Dissecting the role of initial collision geometry for jet quenching observables in relativistic heavy ion collisions}
\newcommand{\sunysb}{Department of Chemistry, Stony Brook University, Stony Brook, NY 11794, USA}
\newcommand{\bnl}{Physics Department, Brookhaven National Laboratory, Upton, NY 11796, USA}
\author{Jiangyong Jia(\CJKfamily{gbsn}¼Ö½­Ó¿)}\email[Correspond to ]{jjia@bnl.gov}
\affiliation{\sunysb}\affiliation{\bnl}
\author{Rui Wei(\CJKfamily{gbsn}κî£)}\affiliation{\sunysb}
\date{\today}
\begin{abstract}
The observation of large azimuthal anisotropy or $v_2$ for hadrons
above $p_T>5$ GeV/$c$ in Au+Au collisions at $\sqrt{s_{\rm
nn}}=200$ GeV has been a longstanding challenge for jet quenching
models based on perturbative QCD (pQCD). Using a simple jet
absorption model, we seek to clarify the situation by exploring in
detail how the calculated $v_2$ varies with choices of the
collision geometry as well as choices of the path length dependence
and thermalization time $\tau_0$ in the energy loss formula.
Besides the change of eccentricity due to distortion from gluon
saturation or event-by-event fluctuation, we find that the $v_2$ is
also sensitive to the centrality dependence of multiplicity and the
relative size between the matter profile and the jet profile. We
find that the $v_2$ calculated for the naive quadratic path length
dependence of energy loss, even including eccentricity fluctuation
and the gluon saturation, is not enough to describe the
experimental data at high $p_T$ ($\sim$ 6 GeV/$c$) in Au+Au
collisions. However, it can match the full centrality dependence of
$v_2$ data if higher power path length dependence of energy loss is
allowed. We also find that the calculated $v_2$ is sensitive to the
assumption of the early time dynamics but generally increases with
$\tau_0$, opposite to what one expects for elliptic flow. This
study attests to the importance of confining the initial geometry,
possibly by combining jet quenching $v_2$ with elliptic flow and
other jet quenching observables, for proper interpretation of the
experimental data.
\end{abstract}
 \maketitle
\end{CJK*}

\section{Introduction}
\label{sec:1} After the discovery of strongly interacting Quark
Gluon Plasma (sQGP) at the Relativistic Heavy Ion Collider (RHIC)
in 2005~\cite{RHIC}, the focus of the heavy ion community shifted
toward a detailed characterization of the properties of the sQGP.
One of the primary tools is jet quenching or the suppression of
high transverse momentum ($p_T$) hadron yields as a result of
in-medium radiative energy loss of high $p_T$
jets~\cite{Gyulassy:2003mc,Wiedemann:2009sh,Majumder:2010qh}. Due
to the large momentum scale of the jets and asymptotic freedom of
Quantum Chromodynamics (QCD), jet quenching is usually thought to
be described by the pertubative QCD (pQCD) framework, which assumes
that jets couple weakly with the medium, even though the medium
itself is strongly coupled. Jet quenching models based on pQCD have
been developed to describe measurements on single hadron
yield~\cite{Adcox:2001jp,Vitev:2002pf}, di-hadron
correlation~\cite{Adare:2008cqb,Zhang:2007ja}, and $\gamma$-hadron
correlation~\cite{Adare:2009vd,Abelev:2009gu,Zhang:2009rn,Qin:2009bk}.
Initial estimates of the properties of sQGP, such as the momentum
broadening per mean free path, $\hat{q}=\langle
k_T^2\rangle/\lambda$, and energy loss per unit length, $dE/dl$,
have been obtained~\cite{Bass:2008rv}.

\begin{figure}[h]
\centering
\epsfig{file=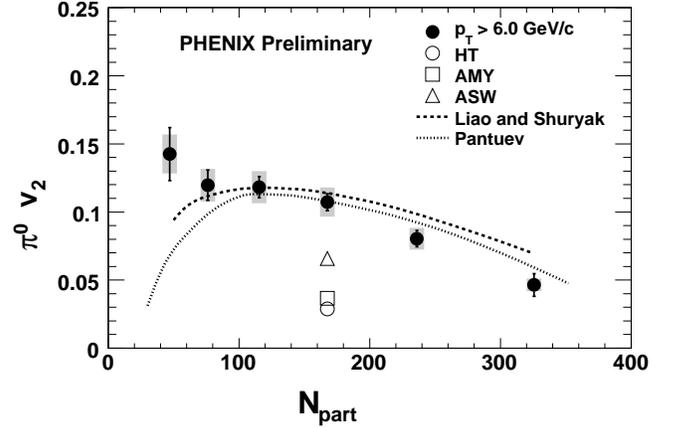,width=1\linewidth}
\caption{\label{fig:0} Figure adapted from Ref.~\cite{Wei:2009mj}.
Data points: PHENIX $\pi^0$ $v_2$ at $p_T>6$ GeV/$c$;
open symbols: three pQCD model calculations taken
from Ref.~\cite{Bass:2008rv}; lines: geometric model calculations
with different assumptions on path-length dependence
~\cite{Pantuev:2005jt,Liao:2008dk}.}
\end{figure}

Despite its early successes, the pQCD description of jet quenching
faces several challenges (see Ref~\cite{Muller:2008zzm}). One
observable that has thus far defied the pQCD description is high
$p_T$ $v_2$ or azimuthal anisotropy of particles emitted relative
to the reaction plane (RP) in Au+Au collisions,
$dN/d(\phi-\Psi_{\rm RP}) \propto (1+2v_2\cos2(\phi-\Psi_{\rm
RP}))$. Such azimuthal anisotropy ensues because the jet yield is
more suppressed along the long axis of the fireball (out-of-plane)
than the short axis (in-plane). Thus, the $v_2$ value is sensitive
to the path length ($l$) dependence of energy loss, which scales,
in the pQCD framework, as $\Delta E\propto l$ and $\Delta E\propto
l^2$ for elastic and Landau-Pomeanchuk-Migdal(LPM) radiative energy
loss~\cite{Peigne:2008wu}, respectively. Currently, most pQCD
models undershoot the $v_2$ value by as much as factor of 2 in the
experimentally accessible $p_T$ range ($p_T<10$
GeV/$c$)~\cite{Shuryak:2001me,Drees:2003zh}. We illustrate this
situation with Fig.~\ref{fig:0} borrowed from
Ref.~\cite{Wei:2009mj}, which compares three mainstream pQCD model
calculations (abbreviated as AMY, ASW, and HT)~\cite{Bass:2008rv}
with recent precision PHENIX data at $p_T\sim 6$ GeV/$c$.

This, together with its failure in describing heavy flavor
suppression~\cite{Adare:2006nq}, call into question the
perturbative assumption used in the pQCD framework. It may happen
that the coupling between the jet and the medium for typical RHIC
jet energy of $p_T\lesssim$ 10-20GeV/$c$ is still strong
enough~\cite{Horowitz:2007su}, such that path length dependence and
the color charge dependence are modified from pQCD expectation. In
fact, calculation based on anti-de Sitter/conformal field theory
(AdS/CFT) technique for strongly coupled plasma suggests that
$\Delta E \propto l^3$~\cite{Gubser:2008as,Dominguez:2008vd} and
$\hat{q} \propto \sqrt{\alpha_{\rm SYM}N_c}$~\cite{Liu:2006ug},
instead of $\Delta E \propto l^2$ and $\hat{q} \propto
\alpha_{s}N_c^2$ for pQCD. This higher order path length dependence
could explain the large anisotropy~\cite{Marquet:2009eq}. Liao and
Shuryak~\cite{Liao:2008dk} argue that most energy loss in sQGP is
concentrated around $T_c$; such a non-monotonic dependence of
energy loss with energy density apparently achieves better
description of the data, as shown by Fig.~\ref{fig:0}.

It is tempting to conclude from this discussion that the data favor
a $l$ dependence stronger than the naive $\Delta E\propto l^2$
implied by the pQCD radiative energy loss. However, as was pointed
out in Ref.~\cite{Drees:2003zh}, the magnitude of the anisotropy is
also very sensitive to the choice of initial collision geometry,
which is poorly constrained. The collision geometry used by most
jet quenching calculations is obtained from the so-called Optical
Glauber model~\cite{Miller:2007ri}, which assumes a smooth
Woods-Saxon nuclear geometry for Au ions. It ignores two important
modifications: an event-by-event distortion of the shape of the
overlap from random fluctuation of positions of participating
nucleons~\cite{Alver:2008zza}; and a possible overall distortion of
the shape of the overlap due to, {\it e.g.}~gluon saturation effect
(so called CGC geometry~\cite{Drescher:2006pi}). Both effects are
shown to lead to 15\%-30\% corrections in the hydrodynamic
calculation of elliptic flow at low $p_T$~\cite{Hirano:2009ah};
they were also shown in
Refs.~\cite{Drescher:2006pi,Broniowski:2007ft} to play an important
role for jet quenching calculation of azimuthal anisotropy at high
$p_T$.

Furthermore, the way that collision geometry influence the jet
quenching $v_2$ is quite different from that for hydrodynamic
description of low $p_T$ $v_2$. Hydrodynamic flow is a self
generating process driven by the shape or eccentricity
($\epsilon=\frac{\langle y^2\rangle-\langle x^2\rangle}{\langle
y^2\rangle+\langle x^2\rangle}$) of a single matter profile, i.e.
$v_2 \propto \epsilon$; whereas the $v_2$ from jet quenching
requires both the profile for the bulk matter AND the jet
production points. The two profiles may not necessarily have the
same spatial distribution because various nuclear effects at
initial state may induce sizable momentum (e.g., Bjorken momentum
fraction $2p_T/\sqrt{s}$) and position dependent modification,
analogous to the generalized parton distribution for proton. Hence
high $p_T$ $v_2$ depends not only on the eccentricity of the
fireball, but also on the matching (relative size and shape)
between the jet and the matter profiles. Understanding the role of
geometry and scaling behavior of the data such as those in
Ref.~\cite{Afanasiev:2009iv,Lacey:2009kg} is important for proper
interpretation of the experimental data.

In this article, we investigate the sensitivity of the jet
quenching $v_2$ on the choices and uncertainties of the collision
geometry for the bulk matter. We check explicitly the scaling and
violation thereof with the bulk eccentricity. We explore, in the
context of these uncertainties, whether the data allow for high
order $l$ dependence of energy loss. The prospects of constraining
the initial collision geometry using $v_2$ and other jet quenching
observables, such as single inclusive suppression $R_{\rm AA}$,
inclusive away-side suppression $I_{\rm AA}$ and associated
anisotropy $v_2^{I_{\rm AA}}$, are discussed.

\section{Model implementation}
\label{sec:2} We generate the Glauber geometry using an improved
version of the publicly available PHOBOS code~\cite{Alver:2008aq}.
Each Au ion is populated randomly with nucleons with a hard-core of
0.3 fm in radii, according to the Woods-Saxon distribution with a
radius of 6.38 fm and diffuseness of 0.535 fm. A nucleon-nucleon
collision is considered to happen when their distance in the
$xy$-plane fall within $\sqrt{\sigma_{\rm nn}^{\rm inel}/\pi}=1.16$
fm (hard-sphere assumption), corresponding a $n$-$n$ cross-section
of $\sigma_{\rm nn}^{\rm inel}$=42 mb. Subsequently, the number
density of nucleons participating in the collision ($\rho_{\rm
part}(x,y,b)$) and the number density of binary collisions
($\rho_{\rm coll}(x,y,b)$) can be determined in the $xy$-plane as
function of impact parameter $b$. Here the $x$ direction is always
chosen to be along the line connecting the centers of the two ions.
Denoting $T_{A}$ as the thickness function for Au ion, they can be
approximated with the following
expression when nucleon size is ignored.
\begin{eqnarray}
\label{eq:1}
\rho_{\rm part}(x,y,b)&\approx&T_A\left(x+\frac{b}{2},y\right)
\left[1-P\left(x+\frac{b}{2},y\right)\right]\nonumber\\
&+&T_A\left(x-\frac{b}{2},y\right) \left[1-P\left(x+\frac{b}{2},y\right)\right],\nonumber\\
\rho_{\rm coll}(x,y,b)&\approx&\sigma_{\rm nn}^{\rm inel} T_A\left(x+\frac{b}{2},y\right)
T_A\left(x-\frac{b}{2},y\right),
\end{eqnarray}
where $P(x,y)=\left(1-\frac{\sigma_{\rm nn}^{\rm inel}
T_A\left(x,y\right)}{A}\right)^A$ and $A=197$ is the number of
nucleons in Au ion.

We generate the CGC geometry using the MC-KLN model by Drescher \&
Nara~\cite{Drescher:2006pi,Drescher:2006ca}, which is based on the
well known KLN (Kharzeev-Levin-Nardi) $k_T$ factorization
approach~\cite{Kharzeev:2002ei}. In a nutshell, the MC-KLN model
calculates the CGC geometry event by event by modifying the output
from a Monte-Carlo Glauber model. Specifically, the transverse
gluon density profile, $dn/dy(x,y,b)$, is calculated through the
$k_T$ factorization formula, with the saturation scale $Q_s^2$ of
each Au ion set to be proportional to its thickness function $T_A$
or $T_{B}$. To ensure internal consistency, the MC-KLN code is
adapted to the same Glauber algorithms as the PHOBOS code (same
hard-core nucleons and identical Woods-Saxon parameters). The
obtained gluon density scales approximately~\cite{Drescher:2006pi}
as $\min\{T_A, T_B\}$ in the $x$ direction and $1/2(T_A+T_B)$ in
the $y$ direction, which leads to a 20\%-30\% increase of the
eccentricity relative to the Glauber geometry (see left panels of
Fig.~\ref{fig:3a}). When implemented in hydrodynamic model
calculations~\cite{Hirano:2009ah,Heinz:2009cv}, a similar amount of
increase is seen for the predicted elliptic flow signal.

We account for initial geometry fluctuation in Glauber geometry by
re-centering and rotating all participants, such that the
``participant plane'' (PP), defined as the minor axis direction of
all participants (see Ref.~\cite{Alver:2008zza}), aligns with the
lab frame. We then sum all events together to give the overall
participant density profile. The same amount of shift and rotation
is then applied for all binary collisions to get the overall
density profile for jet production points. We repeat the same
procedure to the gluon density profile for CGC geometry. Important
variables include the orientation of the participant plane
($\Psi_{\rm part}$) for either participants or gluon
density~\footnote{We emphasize that the PP (and $\Psi_{\rm part}$)
is calculated in the position space of nucleons or gluons in a
simulated collision, whereas experiments measure the so called
event plane (EP) using final state particles in momentum space of a
real collision. Both PP and EP include fluctuations and approximate
the true RP of their respective collisions, but they may not
coincide with each other.}, eccentricity with respect to the
reaction plane ($\epsilon_{\rm RP}$), eccentricity with respect to
the participant plane ($\epsilon_{\rm part}$), and average root
mean square (RMS) size of the ellipsoid ($\sigma_r$). They are
calculated for each event as
\begin{eqnarray}
\nonumber
\tan(2\Psi_{\rm part}) &=&\frac{\sigma_y^2-\sigma_x^2}{2\sigma_{xy}^2},\\\nonumber
\epsilon_{\rm RP} &=&\frac{\sigma_y^2-\sigma_x^2}{\sigma_y^2+\sigma_x^2},\\\nonumber
\epsilon_{\rm part} &=&\frac{\sqrt{(\sigma_y^2-\sigma_x^2)^2+4\sigma_{xy}^2}}{\sigma_y^2+\sigma_x^2}=\frac{\sigma_y^{\prime 2}-\sigma_x^{\prime 2}}{\sigma_y^{\prime 2}+\sigma_x^{\prime 2}},\\
\sigma_{r}^2 &=&\sigma_y^2+\sigma_x^2=\sigma_y^{\prime 2}+\sigma_x^{\prime 2},\label{eq:2}
\end{eqnarray}
where $\sigma_x^2$, $\sigma_y^2$ and  $\sigma_{xy}$ are the
event-by-event (co)variances of participant density profile for
Glauber geometry or gluon density profile for CGC geometry,
respectively, and $\sigma_x^{\prime ^2}$ and $\sigma_y^{\prime ^2}$
are variances defined in the rotated frame. We emphasize that the
participant plane angle, $\Psi_{\rm part}$, should be the natural
frame for both hydrodynamic flow and jet quenching. However, it is
tilted by a different amount in the case of the CGC geometry from
the Glauber geometry.

In this work, the magnitude of the jet quenching $v_2$ depends on
the following four control factors:
\begin{enumerate}
\item{\bf Energy loss formula}, including the path length
    dependence, thermalization time etc.
\item {\bf Eccentricity}, including event-by-event fluctuation
    and the shape of collision geometry (e.g. CGC vs Glauber).
\item {\bf Centrality dependence of the total multiplicity}.
    Because the jet quenching strength is fixed in most central
    collisions, if matter density falls faster toward
    peripheral collisions, we expect less suppression and
    smaller $v_2$ in peripheral collisions.
\item {\bf The size of the matter profile relative to the jet
    profile}. If the transverse size of the matter profile is
    smaller than that for the jet profile, more surviving jets
    should originate from the corona region, leading to a
    smaller $v_2$.
\end{enumerate}
Clearly, the collision geometry (items 2-4) plays an essential role
for proper understanding of the energy loss mechanism (item 1). In
contrast to hydrodynamic description of low $p_T$ $v_2$, which
depends only on the eccentricity of the ellipsoid, jet quenching
description of high $p_T$ $v_2$ is sensitive to two more aspects of
the collision geometry (items 3 and 4). The primary goal of this
work is to understand how the jet quenching $v_2$ depends on the
underlying choices of eccentricity, centrality dependence of
multiplicity, and matching between the matter and the jet profile.

\begin{figure}[ht]
\epsfig{file=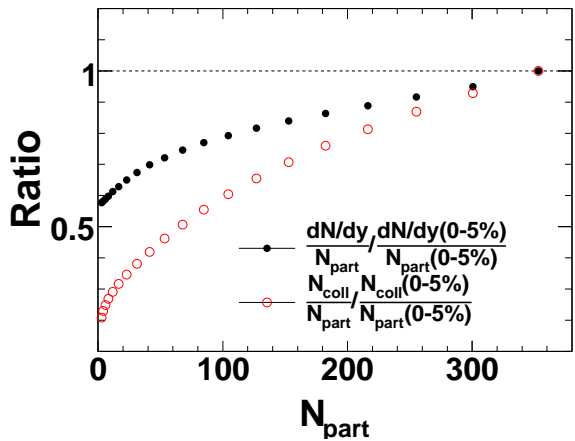,width=0.9\linewidth}
\caption{\label{fig:1a}(Color online) The centrality dependence of $dN/dy/N_{\rm part}$ (filled) and
$N_{\rm coll}/N_{\rm part}$ (open) normalized by their values in most central 0\%-5\% bin.
This plot illustrates different rates of change as a function of centrality, between $N_{\rm part}$,
$N_{\rm coll}$, and $dN/dy$.}
\end{figure}

We base the study on the following four matter profiles (three
versions of Glauber geometry and one CGC geometry)
\begin{eqnarray}
\label{eq:3}\nonumber \rho_{0}(x,y,b) &=& \rho_{\rm part}(x,y,b),\;\;\\\nonumber
\rho_{1}(x,y,b) &=& \rho_{\rm coll}(x,y,b),\;\;\\\nonumber
\rho_{2}(x,y,b) &=& \frac{1-\delta}{2}\rho_{\rm part}(x,y,b)+
\delta\rho_{\rm coll}(x,y,b),\\\nonumber
\rho_{3}(x,y,b) &=& \rho_{\rm CGC}(x,y,b)=dn/dy(x,y,b),\;\;\\\nonumber
\end{eqnarray}
with the corresponding integral form
\begin{eqnarray}
\label{eq:4}\nonumber
\int dxdy \hspace{2mm}\rho_{0}(x,y,b)&=& N_{\rm part} (b), \hspace{6mm}\\\nonumber
\int dxdy \hspace{2mm}\rho_{1}(x,y,b)&=& N_{\rm coll} (b), \hspace{6mm}\\\nonumber
\int dxdy \hspace{2mm}\rho_2(x,y,b) &=& \frac{1-\delta}{2}N_{\rm part}(b)+ \delta N_{\rm coll}(b)\\\nonumber
&=& dN/dy (b),\\\\\nonumber
\int dxdy \hspace{2mm}\rho_{3}(x,y,b)&=&  dN/dy (b),\hspace{3mm}
\end{eqnarray}
where $\rho_{\rm part}$ and $\rho_{\rm coll}$ are transverse
participant density and collision density from the Glauber model,
respectively; $\rho_2$ is the two component Glauber model from
Ref.~\cite{Kharzeev:2000ph} with
$\delta=0.14$~\cite{Hirano:2009ah}, and $\rho_3=\rho_{\rm CGC}$ is
the transverse gluon density from MC-KLN. Both $\rho_2$ and
$\rho_3$ have been adjusted~\cite{Hirano:2009ah} such that their
total integrals match the centrality dependence of the charged
hadron multiplicity, $dN/dy(b)$, at RHIC~\cite{Back:2002uc}. For
the first two profiles, $\rho_0$ and $\rho_1$, we can enforce the
same centrality dependence as $dN/dy(b)$ by applying a centrality
dependent scale factor:
\begin{eqnarray}
\label{eq:5}\nonumber
\rho_{0}^{\rm Mul}(x,y,b) &=& \frac{dN/dy(b)}{N_{\rm
part}(b)}\rho_{\rm part}(x,y,b),\\\nonumber
\rho_{1}^{\rm Mul}(x,y,b) &=& \frac{dN/dy(b)}{N_{\rm coll}(b)}\rho_{\rm coll}(x,y,b),\\
\nonumber
\int dxdy \hspace{2mm} \rho_0^{\rm Mul}(x,y,b) &=&  dN/dy
(b),\\\nonumber
\int dxdy \hspace{2mm} \rho_1^{\rm Mul}(x,y,b) &=&  dN/dy (b),\\
\end{eqnarray}
These scale factors essentially account for the different
centrality dependence trends between $N_{\rm part}$ and $N_{\rm
coll}$ relative to $dN/dy$. Figure~\ref{fig:1a} shows the rate of
change of $dN/dy$ and $N_{\rm coll}$ relative to $N_{\rm part}$
normalized to unity for most central points. Clearly, the $N_{\rm
coll}$ has the fastest change vs. centrality, followed by $dN/dy$,
and $N_{\rm part}$ has the slowest change vs. centrality.
Nevertheless, the resulting profiles, $\rho_0^{\rm Mul}$ and
$\rho_1^{\rm Mul}$, still maintain their original shape and size.
They are used to study the sensitivity of $v_2$ to the centrality
dependence of the multiplicity.

Besides the two profiles obtained by matching to $dN/dy$, we are
also interested in several other variants of $\rho_0-\rho_3$,
obtained either by rotation of their respective participant
planes~\footnote{Note that the participant plane angle $\Psi_{\rm
part}$ is defined separately for $\rho_0$-$\rho_3$.  For $\rho_1$,
it is determined by the minor axis of all binary collisions, while
that for $\rho_2$ is determined by both participants and binary
collision with appropriate weights given in Eq.~\ref{eq:3}.},
\begin{eqnarray}
\nonumber
\rho_n^{\rm Rot}(x,y)&=&\rho_n(x\cos\Psi_{\rm part}-y\sin\Psi_{\rm
part},\\\nonumber
&&x\sin\Psi_{\rm part}+y\cos\Psi_{\rm part}),
\end{eqnarray}
or by readjusting the overall size by a constant scale factor $a$
to match to that of $\rho_0$ (see discussions in Secs.~\ref{sec:3b}
and ~\ref{sec:3c})
\[
\rho_n^{\rm
Resize}(x,y)=\rho_n(ax,ay),
\]
or by a combination of matching multiplicity, rotation or
readjusting size. In the case that a profile is obtained via
several operations, we use appropriate superscript to indicate
that. For example, $\rho_0^{\rm Rot, Mul}$ indicates the matter
profile obtained by rotating the event-by-event participant profile
according to the participant plane angle, followed by matching its
total integral to $dN/dy$ for each centrality bin. Note the order
of these operations has no significance because they factorize.

We implement jet quenching using the simple jet absorption model of
Ref.~\cite{Drees:2003zh}. It provides a transparent way of
investigating the sensitivity of jet quenching observables to
choices of the collision geometry. In this model, back-to-back jet
pairs are generated according to the binary collision density
profile in $xy$-plane with uniform orientation. These jets are then
propagated through the medium whose density is given by matter
profile $\rho(x,y)$, with a survival probability $e^{-\kappa I}$.
In the default setup, matter integral $I$ is calculated as
\begin{eqnarray}\label{eq:6}
I=\int_{0}^\infty dl\frac{l}{l+l_0}
\rho{\left(\overrightarrow{\mathbf{r}}+\left(l+l_0\right)\widehat{\mathbf{v}}\right)}
\approx \int_{0}^\infty dl\hspace{1mm}\rho{\left(\overrightarrow{\mathbf{r}}+l\widehat{\mathbf{v}}\right)}
\end{eqnarray}
for a jet generated at $\overrightarrow{\mathbf{r}}=(x,y)$ and
propagated along direction $\widehat{\mathbf{v}}$. This corresponds
to a quadratic dependence of absorption ($\propto ldl$) in a
longitudinal expanding or 1+1D medium ($\propto 1/(l_0+l)$) with a
thermalization time of $l_0=c\tau_0$. It is fixed to 0 by default
but we explore non-zero value of $l_0$ in Sec.~\ref{sec:3e}. The
absorption coefficient $\kappa$ (which controls the jet quenching
strength) is chosen to reproduce $R_{\rm AA}=\langle e^{-\kappa I}
\rangle\sim0.18$ for 0\%-5\% $\pi^0$ data~\cite{Adare:2008qa}. We
explore path length dependence by extending Eq.~\ref{eq:6} to four
different functional forms,
\begin{eqnarray}
I_m= \int_{0}^\infty dl\hspace{1mm}l^{m-1}
\hspace{1mm}\rho{\left(\overrightarrow{\mathbf{r}}+l\widehat{\mathbf{v}}\right)},\;\;\;\;\; m=1,2,3,4,
\end{eqnarray}
where $m=1$ and $m=2$ correspond to $l$ dependence for radiative,
and AdS/CFT energy loss in 1+1D medium, respectively.

A typical calculation starts by choosing one of the four matter
profiles (Glauber geometry $\rho_0-\rho_2$ or CGC geometry
$\rho_3$) and applying appropriate modifications (specifying Rot,
Mul, and/or Resize). We specify the jet absorption scheme by
varying thermalization time $l_0$ or the order of path length
dependence $m$. We then fix the $\kappa$ value by matching $R_{\rm
AA}\sim 0.18$ in most central collision (We explore the
uncertainties of $\kappa$ arising from experimental uncertainties
of $R_{\rm AA}$, and discuss their implications in
Appendix~\ref{app:AA}). On the other hand, the jet production
profile is always given by $\rho_{\rm coll}$. We stress that
$\kappa$ is the only free parameter, and has similar role as the
$\hat{q}$, and it is tuned {\it independently} for each one of
these running modes (there are $\sim100$ of them, depending on the
choice of matter profiles, $m$, and modifications of matter
profiles). Once $\kappa$ is known, we can predict the centrality
dependence of the single hadron suppression ($R_{\rm AA}$), jet
quenching $v_2$ which can be expressed as $v_2=\langle e^{-\kappa
I} \cos2\left(\phi-\Psi_{\rm part}\right)\rangle$; and away-side
per-trigger yield suppression ($I_{\rm AA}$).

Finally, we point out that the $v_2$-like modulation is found to be
the dominating contribution to the azimuthal anisotropy obtained in
our calculations. The higher order terms, mostly $v_4$, are found
to be less than 10\% of $v_2$ value for all running modes. Thus we
can safely assume that the azimuthal distribution of particle
production relative to the PP angle follows a
$1+2v_2\cos2(\phi-\Psi_{\rm part})$ shape (for example see
Fig.~\ref{fig:8a}).

\section{RESULTS}
\label{sec:3}\subsection{Glauber geometry based on participant
profile} \label{sec:3a}

As mentioned previously, this work investigates three versions of
Glauber geometry, participant profile $\rho_0$, collisional profile
$\rho_1$ and two component profile $\rho_2$ and their variants.
$\rho_0$ is our default Glauber geometry and is the topic of this
section; we shall discuss $\rho_1$ and $\rho_2$ in
Section~\ref{sec:3c}.

\begin{figure}[ht]
\centering
\epsfig{file=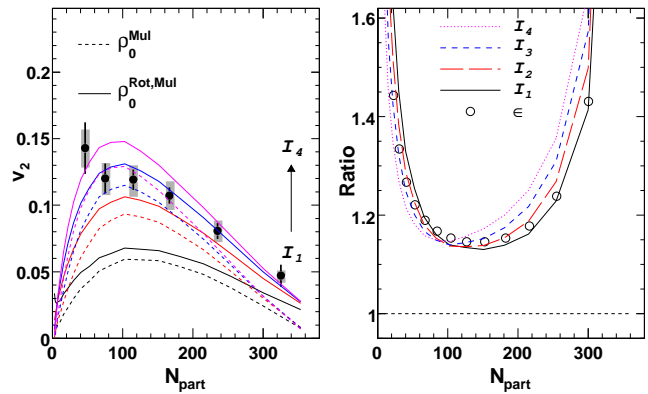,width=1.\linewidth}
\caption{\label{fig:2a} (Color online) Left: $v_2$ calculated with $\rho_0^{\rm Mul}$ (participant density profile scaled to match dN/dy)
with (solid lines) and without (dashed lines) rotation to the participant plane compared with data at 6 GeV/$c$ (solid circles) for $I_1-I_4$ (from bottom up);
Right: Corresponding ratios between with and without rotation for $v_2$s from $I_1-I_4$ (lines) and for the eccentricity (open circles).}
\end{figure}
Figure~\ref{fig:2a} shows the $v_2$ calculated for Glauber geometry
$\rho_0^{\rm Mul}$, that is, participant profile scaled to match
the experimental multiplicity. Results are presented in left panel
for four different path length dependencies ($I_1-I_4$ from bottom
to top) with (solid lines) and without (dashed lines) taking into
account the fluctuation of PP angle. They are compared with the
PHENIX $\pi^0$ $v_2$ data integrated above 6 GeV/$c$ from
Fig.\ref{fig:0}. The right panel shows the ratios of calculated
$v_2$ for $I_1-I_4$ (solid lines) and the ratio of the eccentricity
(open circles) between with and without including the fluctuations.

We see that increasing $m$ (the order of $l$ dependence)
significantly increases the $v_2$ for mid-central collisions, but
they all systematically under-predict the data toward central
collisions. In fact, the calculated $v_2$ for central collision is
insensitive to the functional form of path length dependence, due
to the almost isotropic shape of the overlap. This situation is
dramatically improved when the fluctuation in the PP angle is
included. The relative increase in $v_2$ is about 15\% for
mid-centrality, and is significantly larger for central and
peripheral collisions. This is consistent with previous studies of
low $p_T$ $v_2$ or elliptic flow, which shows that PP fluctuation
needs to be included in Cu+Cu and central Au+Au collisions in order
for hydrodynamic model prediction to
work~\cite{Alver:2008zza,Teaney:2009qa}. It is interesting to see
that the fractional increase of $v_2$ for $I_1$ is similar to the
fractional increase in eccentricity (i.e., $\epsilon_{\rm
part}/\epsilon_{\rm RP}$ in the right panel). However, the ratios
indicate that the fractional increase for $N_{\rm part}>50$ is
successively larger for larger $m$. This is because larger $m$
places more weight to the large $l$ region; and thus is more
sensitive to changes in shape.

\begin{figure}[ht]
\centering
\epsfig{file=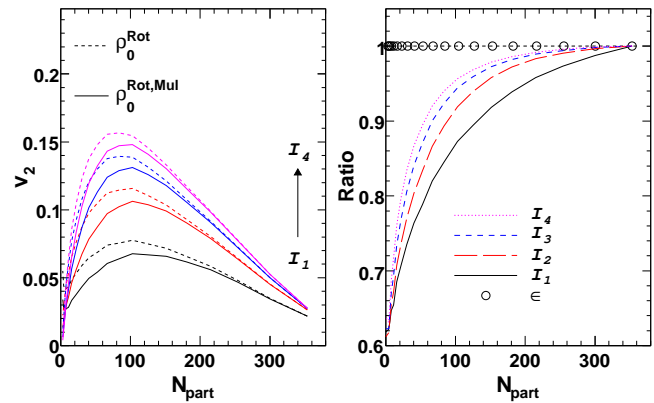,width=1.\linewidth}
\caption{\label{fig:2b} (Color online) Left: $v_2$ calculated with $\rho_0^{\rm Rot}$ (participant density profile rotated to participant plane)
with (solid lines) and without (dashed lines) scaling to match $dN/dy$ for $I_1-I_4$ (from bottom up);
Right: Corresponding ratios between with to without scaling to $dN/dy$ for $v_2$s from $I_1-I_4$ (lines) and for the eccentricity (open circles).
Note that the eccentricities are identical for the two cases.}
\end{figure}

The high $p_T$ azimuthal anisotropy, being the result of jet
quenching, depends not only on the shape, but also on the average
density or total multiplicity of the matter profile. To illustrate
this point, Fig.~\ref{fig:2b} shows the $v_2$ calculated for matter
density $\rho_0^{\rm Rot}$ and $\rho_0^{\rm Rot, Mul}$. They have
identical shape and size for each centrality selection, but the
integral of $\rho_0^{\rm Rot, Mul}$ drops more rapidly to lower
$N_{\rm part}$. Because jet absorption strength $\kappa$ is tuned
to reproduce a common suppression level in central collision, the
profile whose average density varies more rapidly with centrality
is expected to show less suppression and less $v_2$ in peripheral
collisions. Indeed, the calculated $v_2$ for $\rho_0^{\rm Rot,
Mul}$ is smaller than that for $\rho_0^{\rm Rot}$ due to a faster
fall off toward peripheral bin.

Figure~\ref{fig:2b} also shows a weakening of the sensitivity for
larger $m$. This is because the weighting from the large $l$ region
is reduced in peripheral collisions due to a smaller geometrical
size. That reduction is stronger for larger $m$, which leads to
smaller sensitivity for large $m$.

One may argue that since $dN/dy(b)$ is constrained by experimental
data, there should be no uncertainty associated with the modeling
of centrality dependence. However, the matter profiles that were
tuned to Au+Au 200 GeV data typically shows $\sim10$\% deviation
from Cu+Cu or Au+Au at different collision
energies~\cite{Adler:2004zn,Alver:2008ck}. Furthermore, many
current pQCD model calculations use profiles that do not match the
$dN/dy$ data. For example, various 1+1D energy loss models assumes
energy loss or $\hat{q}$ to be proportional to either $\rho_{\rm
part}$~\cite{Zhang:2007ja,Vitev:2005he,Wicks:2005gt} or $\rho_{\rm
coll}$~\cite{Dainese:2004te}; Recent more sophisticated
calculations~\cite{Renk:2006sx,Bass:2008rv,Qin:2009bk} based on
3D+1 hydrodynamics model of Nonaka and Bass~\cite{Nonaka:2006yn},
assume the energy loss or $\hat{q}$ $\propto e^{3/4}$ with
$e\propto 0.6\rho_{\rm coll}+0.4\rho_{\rm
part}$~\cite{Nonaka:2006yn}, which is also different from $dN/dy$.
So it seems reasonable to use the difference of the $v_2$ in right
panel of Fig.~\ref{fig:2b} as one of the uncertainties in
theoretical implementation of initial geometry.

Figure~\ref{fig:2a} and ~\ref{fig:2b} represent the general style
of the presentation of the $v_2$ calculation in this article: The
left panels always shows the $v_2$ values for $I_1-I_4$ compared
between two matter profiles with (solid lines) and without (dashed
lines) a particular geometrical effect; the right panels always
show the ratios between the two (solid line divided by dashed
line). In most cases, the ratio of their eccentricities is shown as
open circles on the right panel to compare with $v_2$ ratios.
Finally, we stress that the jet production profiles are always
sampled from $\rho_{\rm coll}$ throughout this study, so any
difference in the calculated $v_2$ can be attributed to the
differences between the two matter profiles.

\subsection{CGC geometry}
\label{sec:3b} As outlined in the introduction, the MC-KLN model is
built on the standard Monte-Carlo Glauber model. So CGC matter
profile $\rho_3$ contains both the overall modification of shape
due to gluon saturation, and the event-by-event fluctuation
stemming from participant fluctuation at Glauber level. In
addition, we can safely use the same binary collision profile for
the jet production, given that the saturation effects are not
expected to modify hard processes with momentum transfer well above
the saturation scale, $Q^2\gg Q_s^2$.

\begin{figure}[ht]
\centering
\epsfig{file=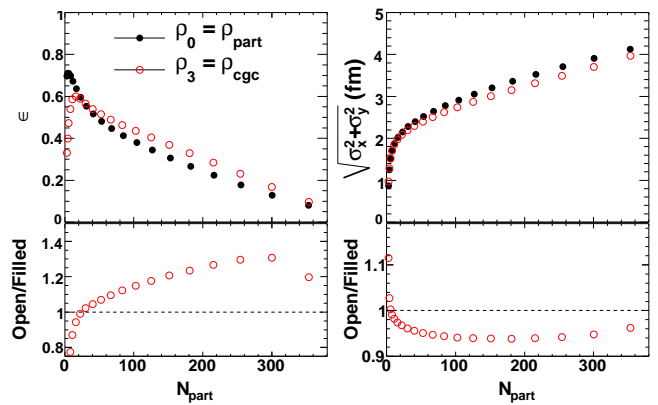,width=1.\linewidth}
\caption{\label{fig:3a} (Color online) Left: Eccentricities for Glauber geometry calculated from participant density profile $\rho_0$ (solid circles)
and for CGC geometry $\rho_3$ (open circles) in the top panel and the corresponding ratio in the bottom panel.
Right: Same as right panels except they are for the RMS size $\sigma_r=\sqrt{\sigma_x^2+\sigma_y^2}$. }
\end{figure}

\begin{figure}[ht]
\centering
\epsfig{file=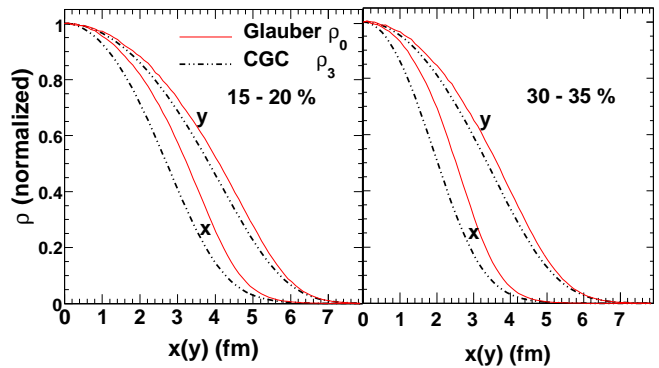,width=1.0\linewidth}
\caption{\label{fig:3b} (Color online)
The Glauber geometry ($\rho_0$) and CGC geometry ($\rho_3$)
projected onto the $x$ and $y$ axes for 15\%-20\% and 30\%-35\% centrality bins.
They are normalized to 1 at the maximum.}
\end{figure}

Another important feature of the CGC geometry via the MC-KLN model
is that despite having a larger eccentricity, its overall size is
about 4\%-8\% smaller than the participant profile. One can see it
quantitatively in Fig.~\ref{fig:3a}, which compare the eccentricity
and overall RMS width ($\sigma_r$) between $\rho_3$ and $\rho_0$.
This narrowing of CGC geometry was pointed out before by the
authors of MC-KLN model (see the preprint version of
~\cite{Drescher:2006pi}), and can be seen more clearly by plotting
the 1D projections of medium profiles along the x (in-plane) and y
(out-of-plane) directions (Fig.~\ref{fig:3b}). The projections show
that CGC profile is narrower than Glauber profile in both x and y
direction, however since $\sigma_x$ is reduced more than $\sigma_y$
in MC-KLN vs Glauber, $\epsilon_{\rm part}$ is larger in MC-KLN.

\begin{figure}[ht]
\centering
\epsfig{file=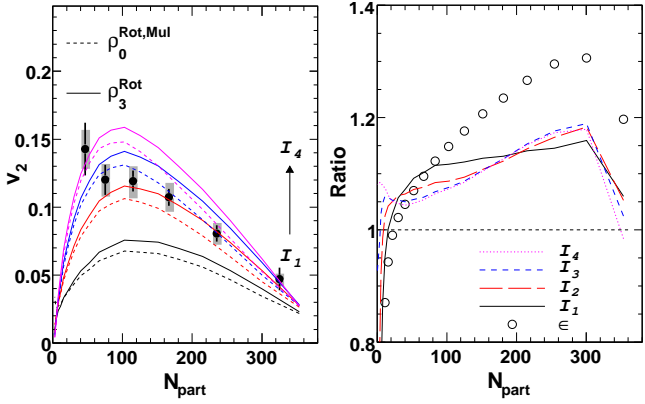,width=1.\linewidth}
\caption{\label{fig:3c} (Color online) Left: $v_2$ calculated for Glauber geometry ($\rho_0^{\rm Rot}$) and CGC geometry ($\rho_3^{\rm Rot}$) in their respective rotated frames for $I_1-I_4$ (from bottom up). Right:
Corresponding ratios for $v_2$s from $I_1-I_4$ (lines) and for the eccentricity (open circles). Note that the CGC geometry has larger eccentricity.}
\end{figure}

\begin{figure}[ht]
\centering
\epsfig{file=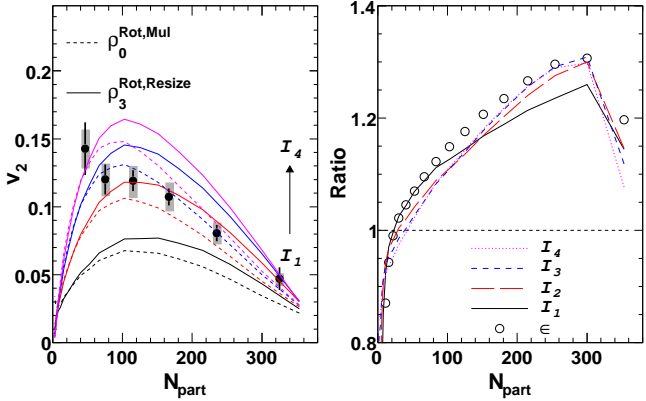,width=1.\linewidth}
\caption{\label{fig:3d} (Color online) Same as Fig.~\ref{fig:3c}, except that the RMS size of CGC geometry
has been stretched to match Glauber geometry $\rho_0$ (according to ratio shown in bottom right panel of Fig.~\ref{fig:3a})}
\end{figure}

Figure~\ref{fig:3c} compares the $v_2$ calculated for CGC geometry
($\rho_3$) and Glauber geometry ($\rho_0$) in their respective
rotated frames. The CGC geometry does lead to a larger $v_2$;
however, the amount of increase is only half of the increase in
eccentricity. To check whether the breaking of the eccentricity
scaling can be attributed to the 4\%-8\% mismatch between the two
profiles, we re-scale the RMS size of the CGC geometry to match
that for the Glauber geometry for each centrality while preserving
its original shape. Figure~\ref{fig:3d} shows the same comparison
after the scaling is applied and $\kappa$ is re-tuned. The ratios
of the calculated $v_2$ now match well with the ratio of the
eccentricities.

These comparison plots clearly show that both eccentricity and the
size contribute to the difference of the $v_2$ between CGC and
Glauber geometry: While the eccentricity of the CGC geometry
increase by about 10\%-30\% relative to Glauber geometry, its
transverse size shrinks. The latter change increases the fractional
contribution of surface jets, which have smaller $v_2$. In
contrast, there is no such bias for hydrodynamic calculation of low
$p_T$ $v_2$, which depends on the shape, not the size, of the
matter profile.

\subsection{Glauber geometry based on collision profile and two component profile}
\label{sec:3c} 
\begin{figure}[ht]
\epsfig{file=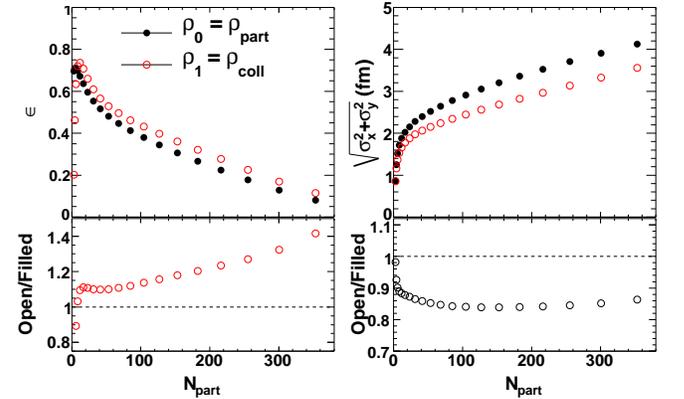,width=1.\linewidth}
\caption{\label{fig:4a} (Color online) Left: Eccentricities for Glauber geometry calculated from participant density profile $\rho_0$ (solid circles)
and from collision density profile $\rho_1$ (open circles) in the top panel and the corresponding ratio in the bottom panel.
Right: Same as right panels except they are for the RMS size $\sigma_r=\sqrt{\sigma_x^2+\sigma_y^2}$.}
\end{figure}

The preceding discussion alludes to an interesting
possibility: For a given energy loss formula and jet production
profile, as long as the matter profile is adjusted to a common
reference $\sigma_r$ and $dN/dy$, the $v_2$ depends only on the
eccentricity of the matter profile. Here we further test this
ansatz by using a matter profile $\rho_1$ that is very different
from $\rho_0$. Comparing to $\rho_0$, $\rho_1$ has much larger
eccentricity (left panels of Fig.~\ref{fig:4a}) which should
increase the calculated $v_2$. On the one hand, it has stronger
centrality dependence of total integral (Fig.~\ref{fig:1a}) and
10\%-15\% smaller $\sigma_r$ (right panels of Fig.~\ref{fig:4a}),
both are expected to significantly decrease the calculated $v_2$.

\begin{figure}[ht]
\centering
\epsfig{file=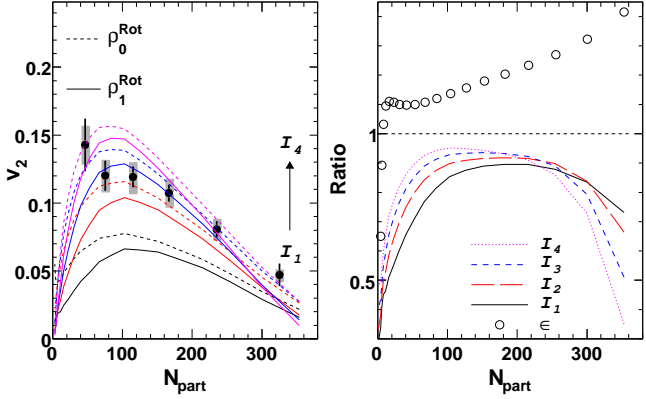,width=1.\linewidth}
\caption{\label{fig:4b} (Color online) Left: $v_2$ calculated for two Glauber geometries, $\rho_0^{\rm Rot}$ and $\rho_1^{\rm Rot}$,
in their respective rotated frames for $I_1-I_4$ (from bottom up). Right:
Corresponding ratios for $v_2$s from $I_1-I_4$ (lines) and for the eccentricity (open circles). Note that $\rho_1^{\rm Rot}$ (collision density profile) has larger eccentricity.}
\end{figure}

Figure~\ref{fig:4b} compares the $v_2$ calculated for $\rho_1$ and
$\rho_0$ in their respective rotated frames. The calculated $v_2$
falls well below the experimental $v_2$ data in central collisions,
which suggests that the Glauber geometry based solely on collision
density profile with eccentricity fluctuation is ruled out. The
decrease of the $v_2$ is largely attributable to the stronger
centrality dependence and smaller size of $\rho_1$, and can be seen
more quantitatively in the right panel, which appears as a large
suppression of the $v_2$ ratios from the expected eccentricity
ratio.

\begin{figure}[ht]
\centering
\epsfig{file=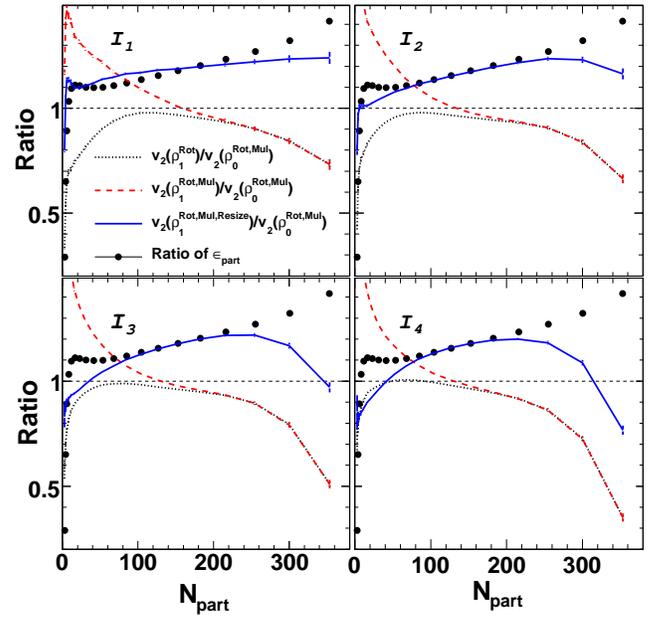,width=1.\linewidth}
\caption{\label{fig:4c} (Color online) Ratios of calculated $v_2$ between $\rho_0^{\rm Rot,Mul}$ and three cases of $\rho_1^{\rm Rot}$:
original multiplicity and size (dotted lines),
multiplicity is adjusted to match $dN/dy$ or $\rho_0^{\rm Rot, Mul}$ (dashed lines),
 both multiplicity and size adjusted to match $\rho_0^{\rm Rot, Mul}$ (solid lines). They are presented separately for $I_1$ (top left panel),
 $I_2$(top right panel), $I_3$(bottom left panel) and  $I_4$(bottom right panel). }
\end{figure}

To dissect the impacts of these factors more clearly, we calculate
the $v_2$ of $\rho_1$ in three different ways before making the
ratio with the $v_2$ of $\rho_0^{\rm Rot,Mul}$: 1) original
multiplicity and size, $\rho_1^{\rm Rot}$; 2) multiplicity is
adjusted to match $dN/dy$ or $\rho_0^{\rm Rot, Mul}$, $\rho_1^{\rm
Rot,Mul}$; 3) both multiplicity and size adjusted to match
$\rho_0^{\rm Rot, Mul}$, $\rho_1^{\rm Rot,Mul,Resize}$. The results
are shown in Fig.~\ref{fig:4c} for $m=1$ (top-left panel), $m=2$
(top-right panel), $m=3$ (bottom-left panel), and $m=4$
(bottom-right panel). Again the $\kappa$ is readjusted
independently for each case. We see that matching the multiplicity
dependence mostly increases $v_2$ at $N_{\rm part}<200$ where the
$dN/dy$ per participant is changing the fastest, but has very
little influence at $N_{\rm part}>200$. However, after the RMS size
of the matter profile is readjusted to match that of the
participant profile, the calculated $v_2$ ratios follow the ratio
of the eccentricities nicely (except in central and peripheral bins
for $m>1$). We clearly see a large sensitivity of $v_2$ on
$\sigma_r$: The large ($\sim50\%$) suppression of $v_2$ in central
collisions, that is, the difference between the dashed line and the
solid circles, is mostly attributable to a $\sim15\%$ narrowing of
$\sigma_r$ (Fig.~\ref{fig:4a}).

\begin{figure}[ht]
\centering
\epsfig{file=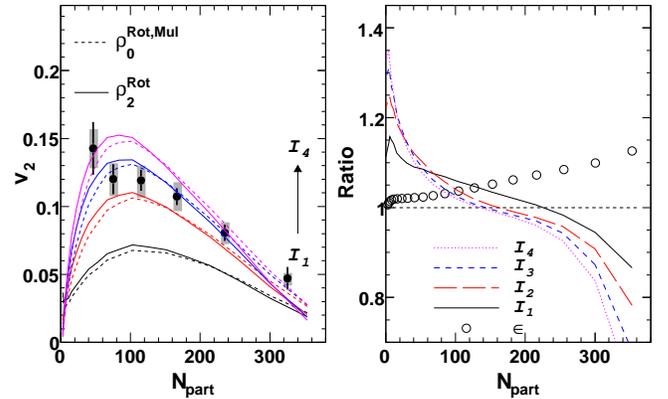,width=1\linewidth}
\caption{\label{fig:4d} (Color online) Left: $v_2$ calculated for two Glauber geometries, $\rho_0^{\rm Rot}$ and $\rho_2^{\rm Rot}$,
in their respective rotated frames for $I_1-I_4$ (from bottom up). Right:
Corresponding ratios for $v_2$s from $I_1-I_4$ (lines) and for the eccentricity (open circles). Note that $\rho_2^{\rm Rot}$ (two component model density profile) has larger eccentricity.}
\end{figure}

It is now straightforward to apply what we learned from
Fig.~\ref{fig:4c} to study the $v_2$ for the two component matter
profile, $\rho_2$. $\rho_2$ is built as a linear combination of
participant density profile ($\rho_1$) and collision density
profile ($\rho_0$). It is a quite popular initial geometry used in
many hydrodynamic model
calculations~\cite{Kolb:2003dz,Chaudhuri:2008sj,Hirano:2009ah}. It
has the correct multiplicity; but a smaller geometrical size
relative to $\rho_0$ due to a centrality-dependent contribution
from binary collision density profile (Even though $\delta$ is
fixed at 0.14, $\rho_{\rm coll}$ becomes more important in central
collisions because $N_{\rm coll}$ grows faster than $N_{\rm part}$
toward central collisions.). We naturally expect the corresponding
eccentricity and $v_2$ should sit between that for $\rho_0$ and
$\rho_1$. This is indeed the case as shown by Fig.~\ref{fig:4d}.
The ratio of the calculated $v_2$ between $\rho_2$ and $\rho_0$ has
shape similar to that of the middle curve in Fig.~\ref{fig:4c},
albeit the rate of change is reduced. The $v_2$ ratios decrease
with $N_{\rm part}$, while the ratio of eccentricity increases
slightly with $N_{\rm part}$. They cross each other at around
$N_{\rm part}\sim100$. In most central collisions the $v_2$ from
two component model is suppressed by about 10\%, even though the
eccentricity value shows $\sim10\%$ increase. Consequently, it
under-predicts the data in central collisions (see left panel).

As a final note, we point out that the mixing fraction,
$\delta=0.14$, between $N_{\rm part}$ and $N_{\rm coll}$ is chosen
to match the $dN/dy$. However, it is not clear that this
parametrization necessarily reflects the true shape and size of the
matter profile (see for example~\cite{Nagle:2009ip}). This concern
is especially true in viewing its poor agreement with the central
data for $I_1-I_4$ even after including the eccentricity
fluctuation. Such poor agreement is clearly due to the narrow
profile of the collision component in the two component profile.
Nevertheless, it seems that by combining the hydrodynamic
description of the low $p_T$ $v_2$ and the jet quenching
description of the high $p_T$ $v_2$, one can gain insight not only
on the eccentricity or shape, but also the size of the matter
profile.
\subsection{Fluctuations beyond rotation to the participant plane}
\label{sec:3d}
\begin{figure*}[ht]
\centering \epsfig{file=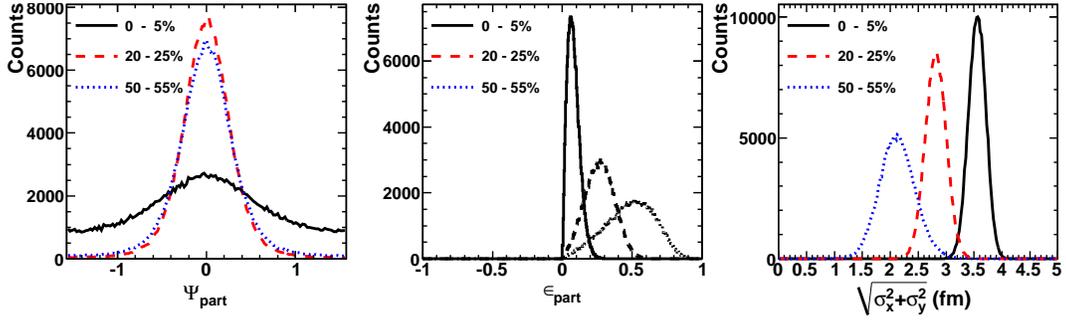,width=0.8\linewidth}
\caption{\label{fig:5e} (Color online) Distributions of participant
plane angle (left panel), participant eccentricity (middle panel)
and RMS size (right panel) for $\rho_0$ in three centrality
selections (0\%-5\%, 20\%-25\% and 50\%-55\%).}
\end{figure*}
The study of fluctuation so far only includes the fluctuation of PP
angle of the initial geometry. In principle, we should also
consider local density fluctuations which affect variance along the
long and short axes of the ellipsoid without changing the
orientation of the PP. In other words, both the size $\sigma_r$ and
the participant eccentricity $\epsilon_{\rm part}$, which are
invariant under rotation, can still fluctuate event to event for
fixed impact parameter. These fluctuations are large compared to
their mean values, as shown by Fig.~\ref{fig:5e}.

Estimation of the influence of these additional fluctuations
requires event-by-event calculation of the jet absorption, where
the nucleons cannot be treated as point-like objects. We assume the
nucleon has a Gaussian profile in the transverse plane with a width
of $r_{0}$ in the $x$ and $y$ directions, corresponding to a
nucleon-nucleon overlap function,
\begin{eqnarray}
\label{eq:7}
t(x,y)=\frac{1}{2\pi r_{0}^2} e^{-\frac{x^2+y^2}{2r_{0}^2}},
\end{eqnarray}
and a binary collision profile,
\begin{eqnarray}
\label{eq:8}
\rho_{\rm coll}(x,y,b)=\int dx^{\prime}dy^{\prime}&&\;
T_{A}(x-\frac{b}{2},y)\times\\
&&T_{A}(x+\frac{b}{2}+x^{\prime},y+y^{\prime})t(x^{\prime},y^{\prime})\nonumber
\end{eqnarray}
The event-by-event participant profile is obtained by summing over
the nucleon profile for all participants.

For each event, we generate four dijet pairs by sampling its
collision profile [Eq.~\ref{eq:8}], then calculate their absorption
in corresponding participant density profile. A total of
$3\times10^6$ Glauber events are used. The $\kappa$ is chosen such
that the overall survival rate averaged over all events is 0.18 for
the 0\%-5\% centrality bin. To check the stability of our result
against the finite size assumption of the nucleons, we varied the
$r_{0}$ from $0.2-0.4$fm, and we also assume the nucleon to be disk
of constant density with a radius of $\sqrt{\sigma_{\rm nn}^{\rm
inel}/\pi}/2=0.58$fm. It turns out the final results are not
sensitive to details of the nucleon overlap function, except for
very peripheral collisions ($N_{\rm part}<20$) when nucleon size
become comparable to the size of the ellipsoid. The deviation is
even smaller, when average collision geometry is used. More
detailed discussion on this can be found in Appendix~\ref{app:B}.

Figure~\ref{fig:5a} shows the influences of these additional
fluctuations on jet quenching $v_2$. Again, the jet absorption
strength is tuned independently to match the central $R_{\rm AA}$
data. The main effect of these additional fluctuations is a small
increase of the $v_2$ in central collisions and a small decrease in
peripheral collisions. The change is less than 10\% for $N_{\rm
part}>100$.

\begin{figure}[ht] \centering
\epsfig{file=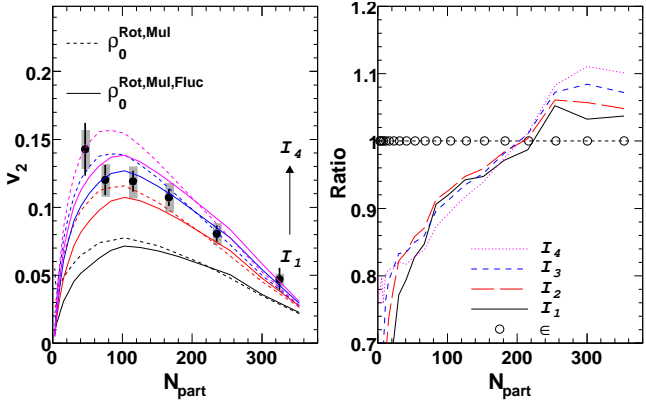,width=1\linewidth}
\caption{\label{fig:5a}(Color online) Left: $v_2$ for participant density profile (but in rotated frame and scaled to $dN/dy$), calculated either on averaged profile ($\rho_0^{\rm Rot, Mul}$)
or event-by-event with additional fluctuations ($\rho_0^{\rm Rot, Mul, Fluc}$). Right:
Corresponding ratios for $v_2$s from $I_1-I_4$ (lines) and for the eccentricity (open circles). Note that
the two cases have the same eccentricities.}
\end{figure}
\begin{figure*}[ht]
\centering \epsfig{file=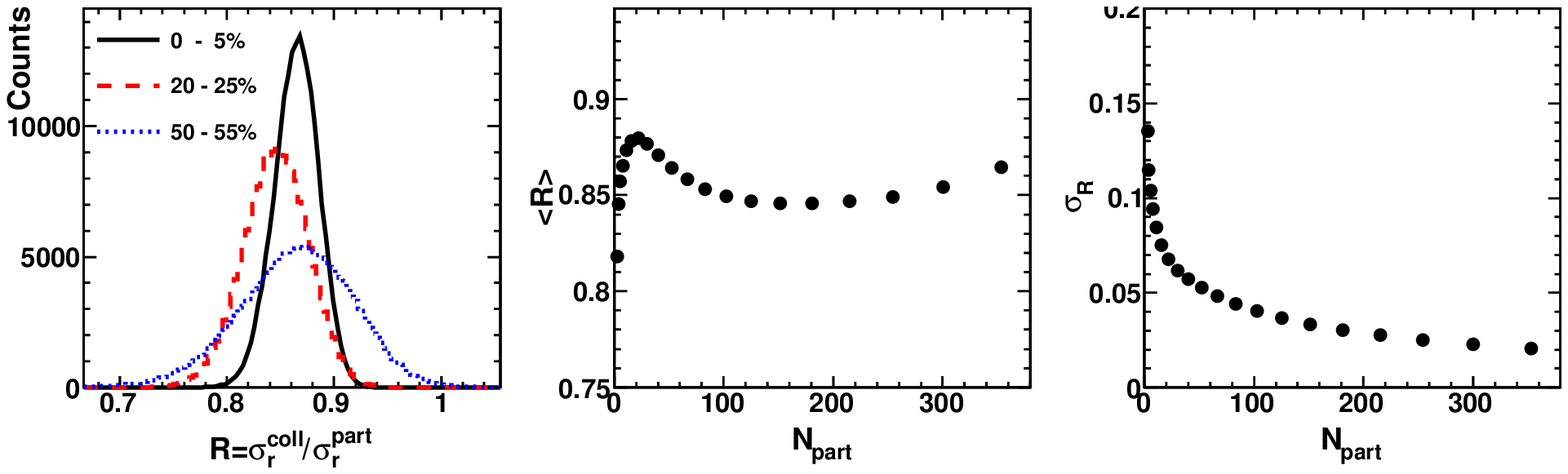,width=0.8\linewidth}
\caption{\label{fig:5d} (Color online) Left: Distributions of
event-by-event ratio of RMS width of collision density profile to
participant density profile, $R = \sigma_r^{\rm coll}/\sigma_r^{\rm
part}$. Middle: centrality dependence of the mean value for $R$,
$\langle R\rangle$. Right: centrality dependence of the RMS width
for $R$, $\sigma_R$.}
\end{figure*}

We can understand this 10\% centrality dependence change of the jet
quenching $v_2$ based on what was learned from previous discussion
as follows. First, we notice that the eccentricity averaged over
many events is not the same as the eccentricity of the matter
profile averaged over many events, that is,
$\left<\frac{\sigma_y^{\prime 2}-\sigma_x^{\prime
2}}{\sigma_y^{\prime 2}+\sigma_x^{\prime 2}}\right> \neq
\frac{\langle\sigma_y^{\prime 2}\rangle-\langle\sigma_x^{\prime
2}\rangle}{\langle\sigma_y^{\prime
2}\rangle+\langle\sigma_x^{\prime 2}\rangle}$. However,
Fig~\ref{fig:A3} in Appendix~\ref{app:B} shows that the difference
is only about 2\% and independent of $N_{\rm part}$ and cannot
explain the difference of the $v_2$ in Fig.~\ref{fig:5a}. Thus, it
must be related to the event-by-event fluctuation of the
$\sigma_r$. What really matters is the relative size between the
jet profile and the matter profile, $R = \sigma_r^{\rm
coll}/\sigma_r^{\rm part}$. Smaller $R$ implies that more jets are
produced in the interior of the matter profile, thus suffer more
energy loss and have large $v_2$; larger $R$ implies more jets are
produced in the corona region of the matter profile, thus suffer
less energy loss and have smaller $v_2$.

Figure~\ref{fig:5d} shows the distribution of $R$ for several
centralities, as well as its mean value $\left<R\right>$ and
standard deviation $\sigma_R$ as a function of centrality.
$\left<R\right>$ is almost constant as a function of $N_{\rm part}$
around $\sim0.85$. However the width of the distribution $\sigma_R$
is a strong function of centrality; it increases from about 2\% in
most central collisions to more than 5\% around $N_{\rm
part}\sim50$, and the distribution becomes asymmetric toward
peripheral bins. It is true that the initial jet production rate
does not depend on the fluctuation of $R$. However, the survival
probability does; that is, more jets escape the medium when $R$
fluctuates to large values while fewer jets escape when $R$
fluctuates to small values. Thus this $R$ dependent survival
probability amplifies the upward fluctuation of $R$, leading to a
smaller $v_2$. The suppression of the $v_2$ is greater in
peripheral collisions due to a broader $R$ distribution. This
explains the falling of the ratio toward peripheral collisions in
Fig.~\ref{fig:5a}.

Before closing this section, we stress that the effect of
fluctuation on $v_2$ can be largely attributed to the fluctuation
of the PP angle. The residual effects, arising mainly from
event-by-event fluctuation of the relative size between matter
profile and jet production profile, are less than 10\% for $N_{\rm
part}>100$. Thus, it seems reasonable to use an averaged matter
profile and binary collision profile for jet quenching calculation,
supplemented with a small centrality dependent correction. This is
of practical importance, because event-by-event jet quenching
calculation is either not possible or computationally prohibitive
for many current pQCD models. However, the lumpiness for
event-by-event geometry implies large fluctuation of scatter
centers along the jet trajectory, which may influence the
LPM~\cite{Migdal:1956tc} coherent effect; there is no such problem
when the event average density profile is used (see
Fig.~\ref{fig:A2}; compare the first three panels with the
bottom-right panel). However, investigation of such effects is
beyond the scope of this study.
\subsection{Dependence on the thermalization time}
\label{sec:3e} One of the main uncertainties in hydrodynamic
description of the elliptic flow arises from modeling of the
thermalization time $\tau_0$, {\it i.e.}~the time at which the
system reaches local equilibrium and hydrodynamic expansion is
turned on. This time also explicitly enters the energy loss
calculation. The value of $\tau_0$ is not known due to lack of
constraints on the initial geometry and early time dynamics. Early
estimation based on ideal hydrodynamics and Glauber
geometry~\cite{Kolb:2000sd} shows that the RHIC $v_2$ data require
$\tau_0<0.6$ fm/$c$ when assuming free-streaming of partons at
$\tau<\tau_0$. However, Luzum and Romatchke~\cite{Luzum:2008cw}
argue that the large initial eccentricity of CGC geometry allows a
bigger $\tau_0$ (up to 1.5 fm/$c$) for free-streaming, without
destroying the agreement of their calculation with experimental
data. Note that free-streaming is an extreme assumption, since
partons always interact with each other and build up flow even if
the matter is not in local thermal
equilibrium~\cite{Vredevoogd:2008id}. This is especially true for
jet energy loss which does not explicitly require local
equilibrium. Nevertheless, it is an interesting question to ask
whether the high $p_T$ anisotropy due to jet quenching can provide
any constraints on $\tau_0$.

Current implementations of the pre-equilibrium energy loss are
different among various pQCD models. The value of $\tau_0$
typically varies in 0-0.6 fm/$c$. Some calculations assume
$\hat{q}=0$, while others assume it is constant at
$\tau<\tau_0$~\cite{Majumder:2007ae,Bass:2008rv,Schenke:2009gb,Majumder:2010qh}.
Although both can describe the single inclusive suppression, the
extracted $\hat{q}$ at $\tau_0$ can differ by as much as a factor
of two~\cite{Armesto:2009zi}. In this work, we tried three
different formalisms to model the pre-equilibrium energy loss.
\begin{itemize}
\item Jets propagate freely to $\tau_0=l_0/c$, then radiative
    energy loss and LPM interference are switched on:
\begin{eqnarray}
\nonumber\label{eq:int1}
I_a&=&\int_{l_0}^\infty dl\hspace{1mm} (l-l_0)
\frac{\rho{\left(\overrightarrow{\mathbf{r}}+l\widehat{\mathbf{v}}\right)}}{l}\\
&=&\int_{0}^\infty dl\hspace{1mm}
\frac{l}{l+l_0}\rho{\left(\overrightarrow{\mathbf{r}}+(l+l_0)\widehat{\mathbf{v}}\right)}
\end{eqnarray}

\item LPM effects start at $\tau=0$, but its contribution to
    energy loss is truncated at $\tau<\tau_0$:
\begin{eqnarray}
\nonumber\label{eq:int2}
I_b&=&\int_{0}^\infty dl\hspace{1mm}
\frac{\rho{\left(\overrightarrow{\mathbf{r}}+l\widehat{\mathbf{v}}\right)}}{l}\times\left\{\begin{array}{ll}
0 &l\leq l_0\\ l &l> l_0 \end{array}\right.
\\&=&\int_{l_0}^\infty
dl\hspace{1mm}
\rho{\left(\overrightarrow{\mathbf{r}}+l\widehat{\mathbf{v}}\right)}
\end{eqnarray}
This functional form is motivated by often made claims similar
    to one from Ref.~\cite{Bass:2008rv}: {\it ``For times prior
to 0.6 fm/$c$, i.e., the starting point of the RFD simulation,
we neglect any medium effects, i.e., assume $\hat{q}$ = 0. Note
that for a purely radiative energy loss model where the average
energy loss grows quadratically with path length in a constant
medium the effect of initial time dynamics is systematically
suppressed and no strong dependence of the energy loss on
variations of the initial time is observed.''}
\item Radiative
energy loss is on all the time, but the experienced density is
        assumed to increase linearly with time and reach the
        local density at $\tau_0$.
\begin{eqnarray}
\label{eq:int3}
I_c&=&\int_{0}^\infty dl\hspace{1mm}\times\left\{\begin{array}{ll}
\rho{\left(\overrightarrow{\mathbf{r}}+l_0\widehat{\mathbf{v}}\right)}\frac{l}{l_0}
&l\leq l_0\\
\rho{\left(\overrightarrow{\mathbf{r}}+l\widehat{\mathbf{v}}\right)} &l> l_0
    \end{array}\right.
\end{eqnarray}
This effectively implies a $\hat{q}$ that grows and reach
        maximum at $\tau_0$.
\end{itemize}

\begin{figure}[ht]
\centering
\epsfig{file=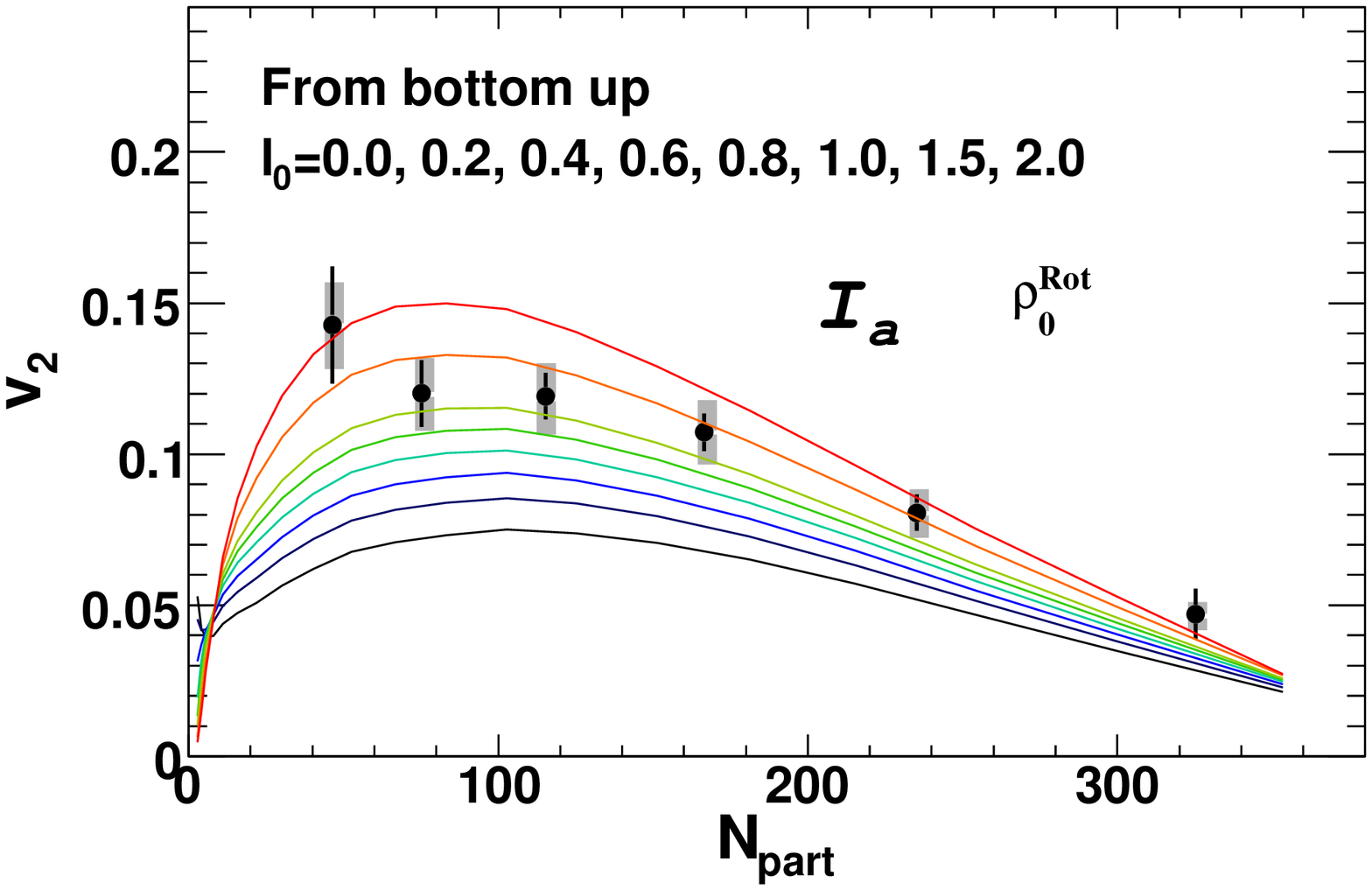,width=0.9\linewidth}
\epsfig{file=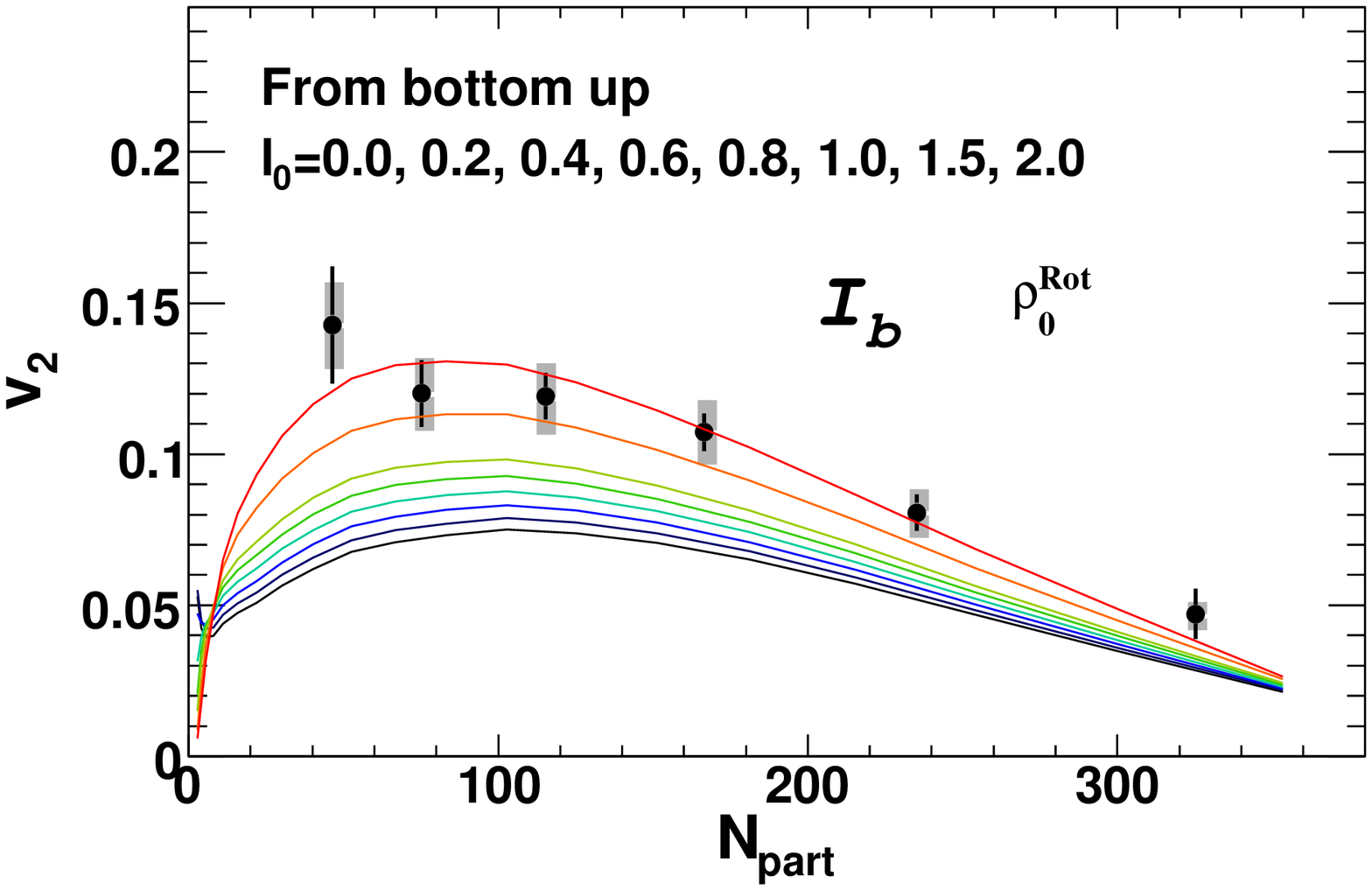,width=0.9\linewidth}
\epsfig{file=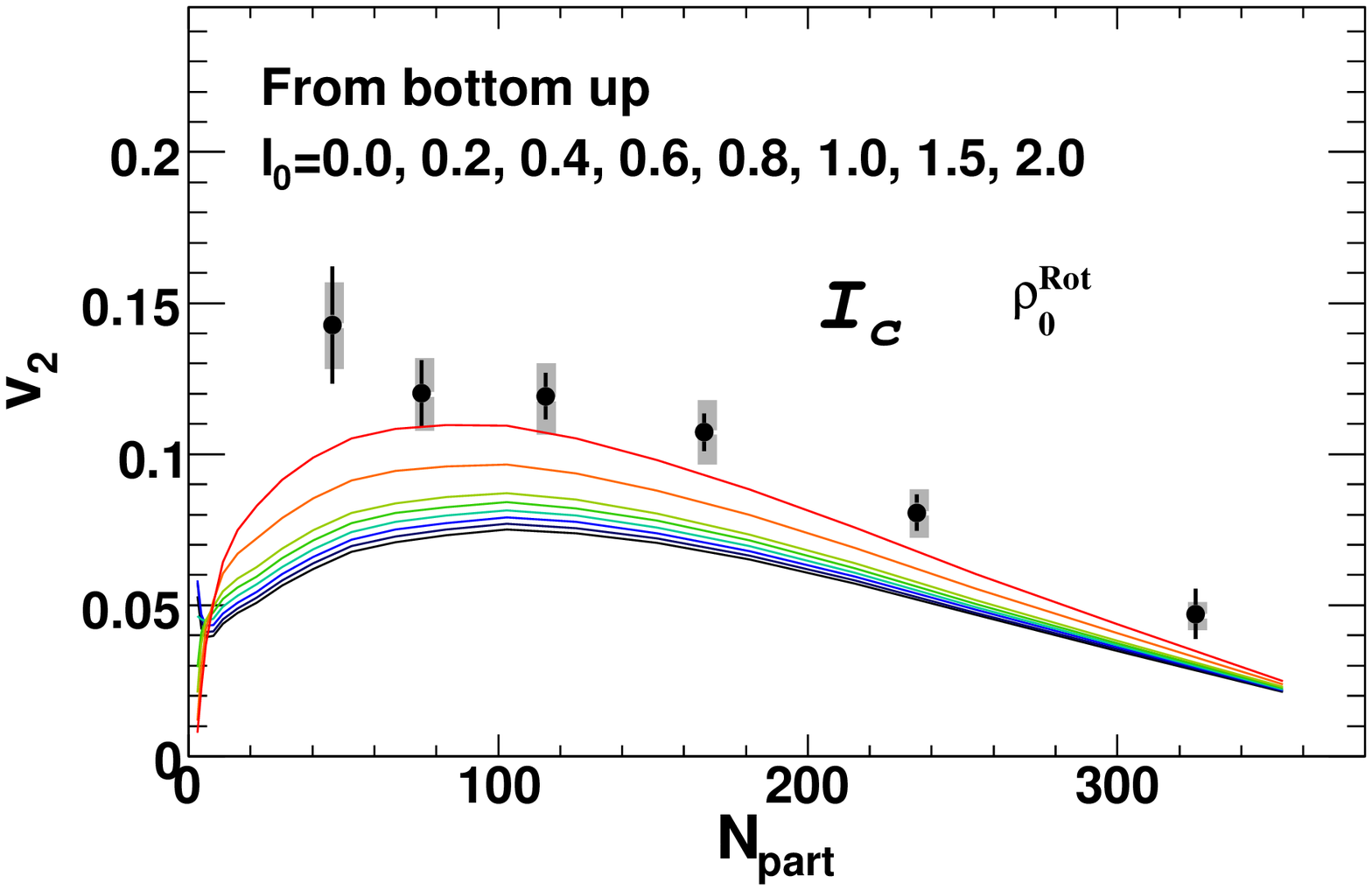,width=0.9\linewidth}
\caption{\label{fig:6a} (Color online) $v_2$ calculated using three different schemes for jet absorptions: $I_a$ according to Eq.~\ref{eq:int1} (top panel),
$I_b$ according to Eq.~\ref{eq:int2} (middle panel), $I_c$ according to Eq.~\ref{eq:int3} (bottom panel). In each cases,
8 different thermalization times range from $l_0$=0 to 2 fm are shown (curves that are lower and darker are for smaller $l_0$).}
\end{figure}

Figure~\ref{fig:6a} shows the calculation for various values of
$\tau_0$ from 0 to 2.0 fm/$c$ for the three cases in the rotated
frame. We find that $I_a$ exhibits the strongest dependence on
$\tau_0$. The calculated $v_2$ increases almost linearly with
$\tau_0$ and reaches the experimental data at $\tau_0\sim$1.5
fm/$c$. This time is somewhat smaller than a similar analysis from
Pantuev~\cite{Pantuev:2005jt}, who need $\tau_0\sim2-3$ fm/$c$ to
match the data. $\tau_0$ is smaller in our case because we include
eccentricity fluctuation. The increase of $v_2$ with $\tau_0$ can
be attributed to increasingly larger contribution from jets
originated from the corona region of the overlap, whose size is
proportional to $\tau_0$~\cite{Pantuev:2005jt}~\footnote{ This can
be qualitatively understood as the following.
Equation~\ref{eq:int1} defines a corona region $l<l_0$ ($I_a$ is
less suppressed) and a core region $l>l_0$ ($I_a$ is more
suppressed), where $l$ is the distance from the surface. The radial
distribution of the initial positions for the surviving jets is
largely defined by the requirement that $R_{AA}=0.18$ or 18\% jets
survive in most central collisions. When $l_0$ is small, the corona
volume is $<18\%$, the surviving jets come from both the corona and
the core and have a quite broad radial distribution. As $l_0$ or
corona volume grows, more and more surviving jets originate from
the corona region with larger anisotropy, and the radial
distribution narrows. Until corona volume reaches about 18\%, most
surviving jets come from corona and the core becomes almost
black.}. Note that the dependence of jet quenching $v_2$ on
$\tau_0$ is just the opposite of that for elliptic flow; the latter
always decreases with increasing $\tau_0$.

For the second functional form, $I_b$, we find that the truncation
of contribution at $\tau<\tau_0$ does simulate the suppression of
the early contribution due to quadratic path length dependence.
However, $I_b$ exhibits a much weaker dependence on $\tau_0$ than
$I_a$, so the two are not equivalent. The increase of $v_2$ reaches
about 20\% for $\tau_0=0.6$ fm/$c$ and grows continuously
thereafter. It reaches the experimental value at $\tau_0=2.0$
fm/$c$ instead of $\tau_0=1.5$ fm/$c$ for $I_a$.

As a more realistic scenario that takes into account some
contributions from $\tau<\tau_0$, $I_c$ exhibits much weaker
dependence on $\tau_0$. The change in $v_2$ is less than 10\% at
$\tau_0<0.6$ fm/$c$; but it again increases quickly at large
$\tau_0$ ($>1$ fm/$c$), where matter integral is dominated by the
corona region which has a large asymmetry.

A slightly different exploration of effects of early time energy
loss has been reported in Ref.~\cite{Armesto:2009zi}. It assumes
either a constant $\hat{q}$ [$\hat{q}=\hat{q}(\tau_0)$] or a
$\hat{q}$ that decreases rapidly to its value at $\tau_0$ [
$\hat{q}=\hat{q}(\tau_0)\left(\frac{\tau_0}{\tau}\right)^{3/4}$].
Both cases imply more pre-equilibrium contribution than $I_c$;
thus, we expect that they have even weaker dependence on $\tau_0$
than on $I_c$. Note that it is generally true that calculation
which has a smaller $\tau_0$ or takes into account the contribution
at $\tau<\tau_0$ always has smaller $v_2$, because the early part
of the matter integral tends to be more isotropic.

Our calculation does not take into account the transverse
expansion. As pointed out earlier~\cite{Lokhtin:2002ux,
Renk:2006pk}, the dependence on $\tau_0$ is further suppressed if
the radial flow is included. This is because the medium moves
outward at speed of $v_T$. Jets that are generated behind the fluid
cell need to move and catch up with it. Thus, matter integrals
decrease more slowly with $\tau_0$ than the 1+1D case. Effectively,
the radial flow tends to shrink the black core region and reduce
the dependence on $\tau_0$.
\subsection{How sensitive is $R_{\rm AA}$ to initial geometry and
energy loss formula?} \label{sec:3f}

The setup of our model framework also provides a convenient way to
study the centrality dependence of several other jet quenching
observables, such as inclusive single hadron suppression $R_{\rm
AA}$, inclusive away-side per-trigger yield suppression $I_{\rm
AA}$ and its associated anisotropy $v_2^{I_{\rm AA}}$ (i.e. $I_{\rm
AA}$ as function of angle with respect to the PP). It is rather
straightforward for us to identify (similar to $v_2$) the most
relevant control factors of the collision geometry and path length
dependence for these observables. To preserve the flow of the main
discussion, we focus this section on the inclusive $R_{\rm AA}$,
because it coupled directly to the $v_2$ discussion. We refer the
reader to some initial work on $I_{\rm AA}$ and $v_2^{I_{\rm AA}}$
in Appendix~\ref{app:A}.

\begin{figure}[ht]
\centering
\epsfig{file=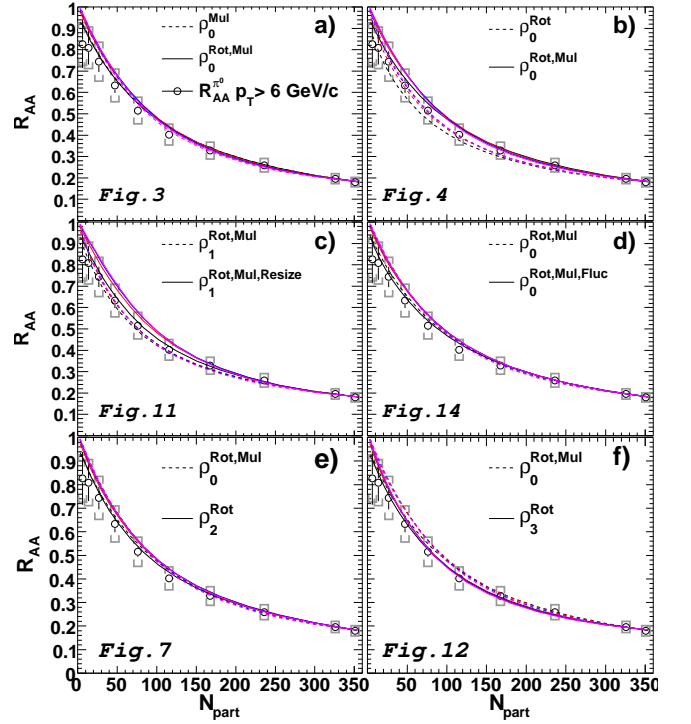,width=1\linewidth}
\caption{\label{fig:7a} (Color online) Each of the six panels show the comparison of centrality dependence of $R_{\rm AA}$ between two matter profiles (matter types are indicated)
 for $I_1-I_4$ (eight curves in total). The corresponding $v_2$ figure for the same set of geometries is indicated in each panel.
 The first four panels shows the effects of switching on and off particular effects of the geometry, i.e. Rotation to participant plane for participant profile (panel a)),
 matching $dN/dy$ for participant profile (panel b)), matching the size for collisional profile (panel c)), and including the additional fluctuation (panel d)).
 The remaining two panels show comparison between Glauber and CGC geometry (Panel e)) and between two Glauber geometries (Panel f)), respectively.}
\end{figure}

\begin{figure}[ht]
\centering
\epsfig{file=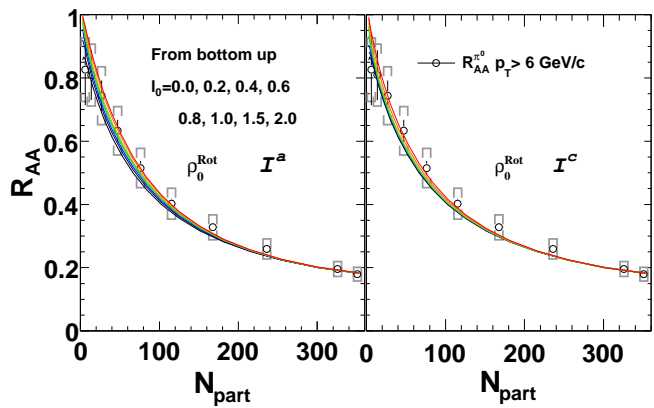,width=1\linewidth}
\caption{\label{fig:7b} (Color online) The centrality dependence of $R_{\rm AA}$ as a function
of thermalization time $l_0/c$. Two different jet absorption schemes are used,
that is, $I_a$ in the left panel (Eq.~\ref{eq:int1}, corresponding $v_2$ figure is
the top panel of Fig.~\ref{fig:6a})
and $I_c$ in the right panel (Eq.~\ref{eq:int3}, corresponding $v_2$ figure is the
top panel of Fig.~\ref{fig:6a}).
Curves that are lower and darker corresponds to smaller $l_0$.}
\end{figure}

Figure~\ref{fig:7a} shows the calculated $R_{\rm AA}$ in six
different cases, with each one designed to check the sensitivity on
one aspect of the geometry. From top to bottom and left to right,
the lists of checked effects as follows:
\begin{itemize}
\item[\bf a)] Without ($\rho_0^{\rm Rot}$) and with
    ($\rho_0^{\rm Rot, Mul}$) eccentricity fluctuation. The
    corresponding $v_2$ comparisons are shown in Fig.~3.

\item[\bf b)] Without ($\rho_0^{\rm Rot}$) and with
    ($\rho_0^{\rm Rot, Mul}$) matching the multiplicity. The
    corresponding $v_2$ comparisons are shown in Fig.~4.

\item[\bf c)] Without ($\rho_1^{\rm Rot,Mul}$) and with
    ($\rho_0^{\rm Rot,Mul,Resize}$) readjusting the RMS size.
    The corresponding $v_2$ are indicated by the dashed and
    solid lines in Fig.~11.

\item[\bf d)] Without ($\rho_1^{\rm Rot,Mul}$) and with
    ($\rho_1^{\rm Rot,Mul,Fluc}$) additional fluctuation beyond
    rotation of the PP. The corresponding $v_2$ comparisons are
    shown in Fig.~15.

\item[\bf e)] Glauber geometry based on ($\rho_0^{\rm Rot,
    Mul}$) versus CGC geometry ($\rho_3^{\rm Rot}$). The
    corresponding $v_2$ comparisons are shown in Fig.~7.

\item[\bf f)] Two glauber geometries: participant profile
    ($\rho_0^{\rm Rot, Mul}$) versus two component profile
    ($\rho_2^{\rm Rot}$). The corresponding $v_2$ comparisons
    are shown in Fig.~12.
\end{itemize}
In each case, all four path-length dependencies $I_1-I_4$ are shown
with and without the particular effect under investigation.
Figure~\ref{fig:7b} compares the different choices of the
thermalization time $\tau_0$ in a broad range using formulation
$I_a$ (left) and $I_c$ (right); the corresponding $v_2$ comparisons
are shown in Fig.~\ref{fig:6a}.

It appears that when jet absorption strength is tuned to reproduce
the suppression in most central collisions, the centrality
dependence of $R_{\rm AA}$ has limited sensitivity on different
choices of the collision geometry and energy-loss formula. This
confirms previous observations~\cite{Drees:2003zh,Eskola:2004cr}
that the centrality dependence of the $R_{\rm AA}$ has limited
discriminating power to dynamics of the underlying energy-loss
mechanisms. This is partly due to the energy loss bias, but it is
also related to the fact that $R_{\rm AA}$ has to vary
monotonically between $R_{\rm AA}\sim0.2$ at large $N_{\rm part}$
and $R_{\rm AA}\sim1$ when $N_{\rm part}\rightarrow0$. However,
$R_{\rm AA}$ does exhibit some sensitivities for three cases; that
is, it increases somewhat for larger $\tau_0$ (Fig.~\ref{fig:7b}),
stronger centrality dependence of $dN/dy$ (Fig.~\ref{fig:7a}b) and
larger RMS size of the matter profile
(Fig.~\ref{fig:7a}c)~\footnote{Naively, one would expect the
$R_{\rm AA}$ to decrease for larger $\sigma_r$ because more jets
originate from inside the profile. However, this is only true if
jet absorption strength $\kappa$ remains the same. Because we
always readjust $\kappa$ such that $R_{\rm AA}\sim 0.18$ in the
most central bin, it almost cancels the expected suppression in
large $N_{\rm part}$ and even makes the peripheral bin less
suppressed (Fig.~\ref{fig:7a}c). This is very different from $v_2$,
which always increases for larger $\sigma_r$ (see
Fig.~\ref{fig:4c}).}. However, the change is well within typical
range of the experimental systematic errors. The situation is very
different for $v_2$ or anisotropy of $R_{\rm AA}$, which is not a
monotonic function of $N_{\rm part}$, and exhibits a much greater
sensitivity to the variation of geometry and energy-loss scheme.

\section{Discussion and Summary}
\label{sec:4}
\begin{table*}
\caption{\label{tab:1} The sensitivity of the $v_2$ and $R_{\rm AA}$ for various changes in matter profile (relative to participant profile)
and energy loss schemes.}

\setstretch{1.2}
  \begin{ruledtabular} \begin{tabular}{l|l|l}
Types of changes               &         $v_2$         & $R_{\rm
AA}$\\\hline \multirow{3}{1.8in}{Fluctuation of RP angle}
&\multirow{3}{3.3in}{Increase by 15\%-30\% for mid-centrality, much
larger in central and peripheral (Fig.~\ref{fig:2a})} &
\multirow{3}{1.6in}{$<5\%$}\\
               &                       &                           \\
               &                       &                           \\\hline
\multirow{3}{1.8in}{Additional fluctuation (relevant for e-b-e jet quenching calculation)}        & \multirow{3}{2.1in}{$\pm 5\%-10$\% (Fig.~\ref{fig:5a})}  &$<5\%$\\
                                                 &                           & \\
                                                 &                           & \\\hline

\multirow{3}{1.8in}{Change in average shape}
&\multirow{3}{3.3in}{Increase by 15\%-30\% for CGC
(Fig.~\ref{fig:3c} and \ref{fig:3d}), by 10\% for two component
model (Fig.~\ref{fig:4d}), 10\%-40\% for $\rho_{\rm coll}$ profile
(Fig.~\ref{fig:4c})} &
\multirow{3}{1.6in}{$<5$\%}\\
                                                 &                           &\\
                                                 &                           &\\\hline

\multirow{3}{1.8in}{RMS size of matter profile ($\sigma_r$)} &
\multirow{3}{3.3in}{$\sim$0\% to -10\% change for CGC (compare
Fig.~\ref{fig:3c} to \ref{fig:3d}), $\sim$10\% to -40\% change for
$\rho_{\rm coll}$ (Fig.~\ref{fig:4c}, compare dashed to solid
line)} &
\multirow{3}{1.6in}{10\% change for 10\% difference}\\
                                                 &                           &\\
                                                 &                           &\\\hline
\multirow{2}{1.8in}{Centrality dependence of multiplicity} & \multirow{2}{3.3in}{Important for $N_{\rm part}<150$ (Fig.~\ref{fig:2b})}&
\multirow{2}{1.6in}{10\%-20\% for mid-centrality}\\
               &                                               &\\\hline
Power $n$ for ``$l$'' dependence  &  Very sensitive & $<5$\%                     \\\hline
\multirow{4}{1.8in}{Thermalization time $\tau_0$}&\multirow{4}{3.3in}{Sensitive but depends on modeling of energy loss at $\tau<\tau_0$.
For $\tau_0=0.6$fm/$c$, it increases by $\sim30$\% for $I_a$, $<15\%$ for $I_b$ and $<5\%$ for $I_c$. (see Fig.~\ref{fig:6a})}  &
\multirow{4}{1.6in}{No, except for $N_{\rm part}<100$}\\
               &                                                  &\\
               &                                                  &\\
               &                                                  &\\\hline
\multirow{2}{1.8in}{12\% variation for $R_{\rm AA}$ in 0\%-5\% central bin}        & \multirow{2}{2.1in}{$ <7$\% at $N_{\rm part}>100$ (Fig.~\ref{fig:AA0})}  &\multirow{2}{1.6in}{$<12\%$}\\
                                                 &                           & \\\hline
\end{tabular}   \end{ruledtabular}

\end{table*}

Table~\ref{tab:1} summarizes the sensitivity of $v_2$ and $R_{\rm
AA}$ on the choices of matter profile and energy-loss formula (also
on experimental uncertainties; see Appendix~\ref{app:AA}). The
former includes fluctuation and shape distortion, which affect the
eccentricity, centrality dependence of multiplicity, and the size
of the matter profile. The latter includes different choices of
path length dependence and thermalization time. Overall, $R_{\rm
AA}$ is not sensitive to these factors, while $v_2$ has fairly
strong dependence on both initial geometry and energy loss formula.
We can summarize the main findings for $v_2$ as follows.
\begin{itemize}
\item Eccentricity fluctuation has to be included in the jet
    quenching calculation. Without it one can not account for
    the large $v_2$ observed in central collisions. We estimate
    that it leads to 15\%-20\% increase of the $v_2$ in the
    mid-centrality bin (Fig.~\ref{fig:2a}).

\item The residual event-by-event fluctuation, other than from
    eccentricity fluctuation, at most leads to 4\%-8\% change
    in $v_2$ at $N_{\rm part}>100$ (Fig.~\ref{fig:5a}).

\item A reasonable variation of the multiplicity dependence,
    only significantly changes the $v_2$ at $N_{\rm part}<100$
    (Fig.~\ref{fig:2b}).

\item CGC geometry always results in a larger $v_2$ than
    Glauber geometry at $N_{\rm part}>150$. Depending on the
    choices of Glauber geometry, the increase ranges anywhere
    from 10\% ($\rho_{\rm part}$, Fig.~\ref{fig:3c}), 30\% (two
    component profile $\rho_{2}$, Fig.~\ref{fig:4d}), or 20\%
    to $>50$\% ($\rho_{\rm coll}$, Fig.~\ref{fig:4c}).

\item $v_2$ is very sensitive to change of the RMS width
    $\sigma_r$ of the profile. A 15\% change in $\sigma_r$ can
    lead to about 30\%-40\% change in calculated $v_2$
    (Fig.~\ref{fig:4c}).
\end{itemize}
These same conclusions seem also to apply for the anisotropy of
away-side suppression $v_2^{I_{\rm AA}}$ (see Fig.~\ref{fig:8c}),
except that the sensitives seem to be much stronger. We stress that
the dependence of jet quenching $v_2$ on geometry is different from
that for the low $p_T$ $v_2$ driven by collective expansion. The
latter is sensitive only to the eccentricity of matter profile
(items 1 and 3 in Table~\ref{tab:1}).

Despite the rather complicated dependence on the initial geometry
for jet quenching $v_2$, most of them, such as the fluctuation and
distortion due to saturation, can be constrained independently by
elliptic flow data~\cite{Luzum:2008cw,Hirano:2009ah}. Jet quenching
$v_2$ appears to be rather sensitive to the choices of energy-loss
formula, thus making it an ideal observable for gaining insights on
energy-loss mechanisms. Based on the comparisons shown in
Fig.~\ref{fig:2a} and \ref{fig:3c}, it seems that naive path length
dependence motivated by radiative energy loss, $I_1$, is
insufficient to describe the data for both Glauber and CGC geometry
even with eccentricity fluctuations taken into account. It appears
that either $2<m<3$ for Glauber geometry based on participant
profile or $m\sim2$ for CGC geometry, both with eccentricity
fluctuations, have the best match with the data. Note that $m=2$
corresponds to the AdS/CFT type of energy loss $\Delta E\propto
l^3$ for a strongly coupled plasma~\cite{Dominguez:2008vd}. Similar
strong path length dependence is also observed for away-side
suppression $I_{\rm AA}$ and it's anisotropous $v_2^{I_{\rm AA}}$
(see Fig.~\ref{fig:8b} and \ref{fig:8c}).

The jet quenching $v_2$ is also quite sensitive to thermalization
time $\tau_0$, but the sensitivity depends on the modeling of the
pre-equilibrium energy loss. By assuming free-streaming up to
$\tau_0=1.5$ fm/$c$ and including eccentricity fluctuation, we can
reproduce the experimental data with quadratic path-length
dependence of energy loss. However, inclusion of a very modest
pre-equilibrium energy loss, for example, $I_c$ which assumes a
$\hat{q}$ which linearly grow to $\hat{q}(\tau_0)$ at $\tau_0$,
already significantly suppressed the dependence on $\tau_0$ up to 1
fm/$c$. We can draw similar conclusions for away-side suppression
$I_{\rm AA}$ and the associated anisotropy $v_2^{I_{\rm AA}}$ (see
Figs.~\ref{fig:8d} and \ref{fig:8e}).

The preceding discussion attested to the value of jet quenching
$v_2$ in understanding the roles of various geometry factors and
constraining the energy loss mechanisms. One can obtain more
discriminating power by combining all jet quenching observables,
$v_2$, $R_{\rm AA}$, $I_{\rm AA}$, and $v_2^{I_{\rm AA}}$ (see
Appendix~\ref{app:A} for more discussions). Initial theoretical
work already demonstrated the value of combining the $R_{\rm AA}$
and $I_{\rm AA}$~\cite{Zhang:2007ja,Armesto:2009zi}; One can do a
better job by also including a calculation of $v_2$. It is worth
pointing out that the study of high $p_T$ $v_2$ benefited
significantly from extensive experimental and theoretical work on
the low $p_T$ $v_2$, which has provided important constraints on
the initial eccentricity. We have shown that the reverse is also
true; {\it i.e.}~jet quenching observables can also provide useful
new insights on the initial geometry, which could help the
interpretation of the elliptic flow data.

In summary, using a simple jet absorption framework, we studied the
sensitivity of jet quenching $v_2$ to various aspects of collision
geometry and the path-length dependence of energy loss. Besides the
eccentricity, we found two other ingredients of the collision
geometry, namely, the centrality dependence of the matter integral
and the relative size between the matter and the jet profiles, are
important for jet quenching $v_2$. We compare the calculated $v_2$
from both Glauber and CGC geometry with experimental data. A path
length dependence stronger than the native $\Delta E\propto l^2$
dependence from radiative energy loss, as well as the inclusion of
the eccentricity fluctuation, are necessary to reproduce the $v_2$
data. A detailed comparison between Glauber and CGC geometry shows
that a 15\%-30\% increase of initial eccentricity in CGC only
results in half the increase in calculated $v_2$ due to a small
narrowing of the CGC geometry. This points to an interesting
possibility: A large $v_2$ can be easily generated if the jet
production profile is significantly narrower than the matter
profile. This requires a transverse profile distribution that
narrows with momentum or Bjorken variable $x=2p_T/\sqrt{s}$. This
happens for the nucleon-nucleon
collisions~\cite{Frankfurt:2003td,Ji:2003ak}; however, we are not
aware yet of a physical mechanism to produce a significant
narrowing at large $x$ in heavy nuclei.

Our estimations are based on a simple jet absorption framework.
Admittedly, it is too simplistic to give direct insight on the
dynamics of the energy-loss process. However, it proves to be a
useful tool for understanding the centrality dependence of various
jet quenching observables, for identifying the most relevant
factors in the collision geometry and path-length dependence, as
well as for estimating the sign and magnitude of the change as we
vary those factors. We have documented all the matter and jet
profiles used in this study in Ref.~\cite{link}. They can be used
as input for future, more-realistic jet quenching calculations.

\section*{Acknowledgement}
We thank W.~Horowitz for stimulating discussions and a careful
proof reading of the manuscript; We appreciate valuable discussions
with R.~Lacey and U.~Heinz; We thank H. J. Drescher, A. Dumitru and
Y. Nara for providing the MC-KLN code. This research is supported
by NSF under award number PHY-0701487 and PHY-1019387.
\appendix
\section{$I_{\rm AA}$ and its azimuthal anisotropy?}
\label{app:A} Initial geometry should also leave footprints on the
azimuthal distribution of the away-side jets. The appropriate
observable for this purpose is the anisotropy of the per-trigger
yield for the away-side jets, $I_{\rm AA} (\phi_{\rm
trig}-\Psi_{\rm part})$, which reflects the path length dependence
of the energy loss for the away-side jet. Due to the surface bias
of the trigger jets, the away-side jets on average have longer path
length to traverse; thus, they are expected to exhibit stronger
suppression and larger anisotropy. In this work, we calculate the
anisotropy coefficient as
\begin{eqnarray}
v_{2}^{I_{\rm AA}} = \langle I_{\rm AA} \cos2(\phi_{\rm trig}-\Psi_{\rm part})\rangle
\end{eqnarray}

\begin{figure}[ht]
\centering
\epsfig{file=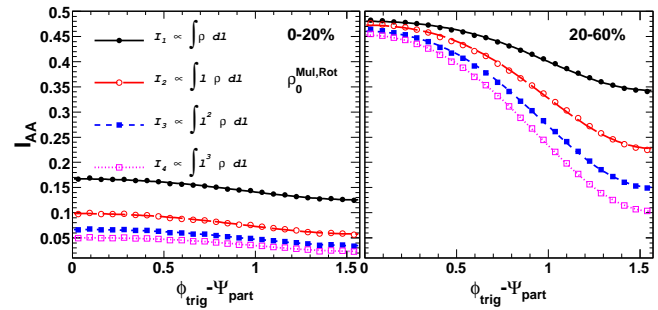,width=1.\linewidth}
\caption{\label{fig:8a} (Color online) The $I_{\rm AA} (\phi_{\rm trig}-\Psi_{\rm part})$ for 0\%-20\% (left panel) and 20\%-60\% (right panel) centrality bins.
The points are results of calculation,
the line is a fit to $1+2v_2^{I_{\rm AA}}\cos2\Delta\phi+2v_4^{I_{\rm AA}}\cos2\Delta\phi$.
We see a finite $v_4^{I_{\rm AA}}$ signal, but it is less than 10\% of $v_2^{I_{\rm AA}}$,
and is ignored in this study.
}
\end{figure}

As a warm-up exercise, Fig.~\ref{fig:8a} shows the azimuthal
dependence of per-trigger yield suppression $I_{\rm AA}$ in
0\%-20\% and 20\%-60\% centrality, calculated for $\rho_0^{\rm
Rot,Mul}$ profile, i.e. participant density profile in rotated
frame and re-scaled to match the $dN/dy$ data. As one increase the
power $m$ of the path length dependence in $I_m$, $I_{\rm AA}$ for
the 20\%-60\% bin shows a dramatic decrease in the out-of-plane
direction where the path length is large, but only a modest
decrease in the in-plane direction where the path length is small.
That is because the suppression for a given centrality is largely
determined by the typical matter integral $I_m\sim
\left<L\right>^m/\left<L_0^m\right>^m$, which changes more rapidly
for larger $m$. Here $L_0^m$ is some typical length scale fixed in
central collisions for $m$. Clearly, the increased sensitivity of
large $m$ can generate a large anisotropy, hence large $v_2^{I_{\rm
AA}}$. However, the price one has to pay is that it leads to a
large suppression in the 0\%-20\% bin as shown in the left panel of
Fig.~\ref{fig:8a}. This is because a large $v_2^{I_{\rm AA}}$
naturally implies a strong suppression in central collisions, as
long as the suppression is a monotonic function of the path length.

The jet absorption framework used in this work is a pure
geometrical model in that the calculated suppression depends only
on the path length, thus it always predicts $I_{\rm AA}<R_{\rm AA}$
due to the longer path length of the away-side jet. However,
experimental data seem to suggest that $I_{\rm AA}\gtrsim R_{\rm
AA}$~\cite{Adams:2006yt,Adare:2008cqb,adare:2010ry}. This is
because that the away-side associated hadron spectra are much
flatter than the inclusive distribution, due to the requirement of
a high $p_T$ trigger. For a typical trigger of 5 GeV/$c$, the
away-side conditional spectra in $p+p$ collisions at $\sqrt{s}=200$
GeV, if parameterized via a power law function $1/p_T^n$, have a
power of $n=8$, in contrast to $n=4$ for inclusive
hadrons~\cite{Jia:2007qi}. A simple estimation shows that to reach
the same level of suppression, it takes about 50\% more energy loss
for away-side jets than for inclusive jets~\cite{Jia:2007qi}. What
this means is that treating energy loss as absorption is not
sufficient; an energy shift term is required as well. However,
phenomenologically, we can still use the jet absorption framework,
if we allow the $\kappa$ to also depend on power $n$; i.e, a
smaller $\kappa$ is required for an away-side jet due to a flatter
input spectra. Such dependence in principle can be fixed by the
$I_{\rm AA}$ data from STAR and
PHENIX~\cite{Adams:2006yt,Adare:2008cqb,adare:2010ry}. However, we
shall defer this improvement to a future study.

\begin{figure}[ht]
\centering
\epsfig{file=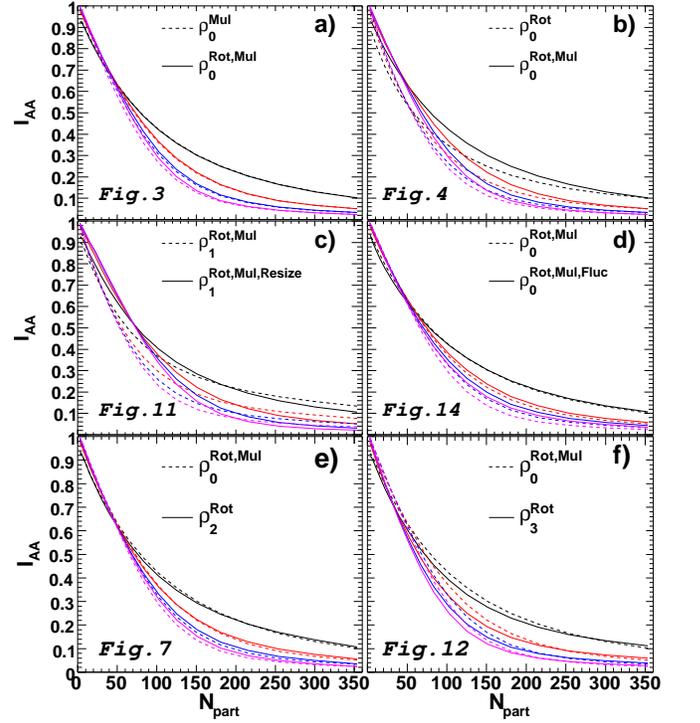,width=1\linewidth}
\caption{\label{fig:8b} (Color online) Conventions are similar to those in Fig.~\ref{fig:7a}.
Each of the six panels shows the comparison of centrality dependence of $I_{\rm AA}$ between two matter profiles (matter types are indicated)
 for $I_1-I_4$ (eight curves in total); the corresponding $v_2$ figures for the same set of geometries are indicated in each panel.
 The first four panels shows the effects of switching on and off particular effects of the geometry, i.e. rotation of participant plane for participant profile (panel a)),
 matching $dN/dy$ for participant profile (panel b)), matching the size for collisional profile (panel c)), and including the additional fluctuation (panel d)).
 The remaining two panels show comparison for Default Glauber vs. CGC geometry (Panel e)) and between two Glauber geometries (Panel f)), respectively.
Note that the small difference in panel a) between
with and without rotation for $m=4$ should be attributed to increased sensitivity
of $I_{\rm AA}$ to changes in $R_{\rm AA}$ for large $m$ (because $\kappa$ is tuned by hand to
match $R_{\rm AA}\sim0.18$ only to the third digit after 0).}
\end{figure}

\begin{figure}[h!]
\centering
\epsfig{file=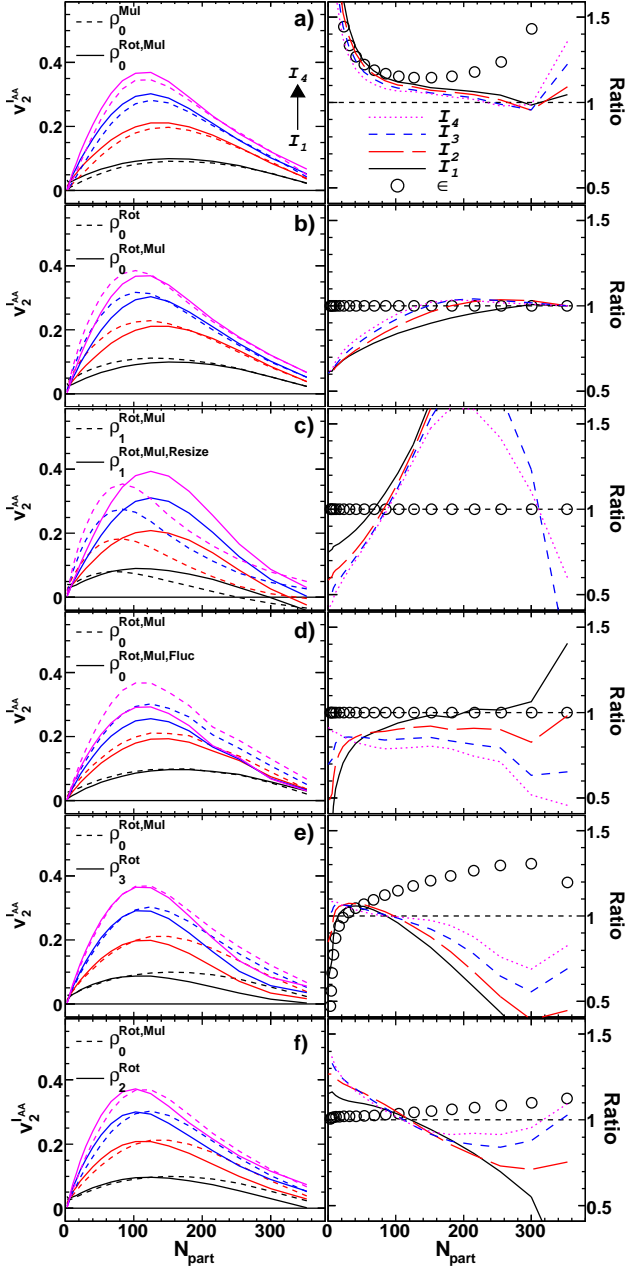,width=0.96\linewidth}
\caption{\label{fig:8c} (Color online) Conventions are similar to those in Fig.~\ref{fig:7a}.
 Each of the six rows show: (left panel) the comparison of centrality dependence of $v_{2}^{I_{\rm AA}}$
between two matter profiles (matter types are indicated), (right panel)
The corresponding ratios together with ratios of eccentricities. The corresponding $v_2^{I_{\rm AA}}$ plot for the same set of geometries is ndicated in each panel.
The first four rows show the effects of switching on and off particular effects of the geometry, i.e. Rotation to participant plane for participant profile (Row a)),
matching $dN/dy$ for participant profile (Row b)), matching the size for collisional profile (Row c)), and including the additional fluctuation (Row d)).
The remaining two rows show comparison for Default Glauber vs. CGC geometry (Row e)) and between two Glauber geometries (Row f)), respectively.}
\end{figure}

Nevertheless, the current setup is sufficient for studying the
sensitivity of $I_{\rm AA}(\Delta\phi)$ on the choices of
underlying collision geometry. Figures~\ref{fig:8b} and
~\ref{fig:8c} summarize the relative change of the inclusive
$I_{\rm AA}$ and its anisotropy $v_2^{I_{\rm AA}}$, respectively,
as one varies various aspects of the collision geometry. The six
cases in both figures correspond to the same change in collision
geometry as in Fig.~\ref{fig:7a} for inclusive $R_{\rm AA}$, that
is, {\bf a)} eccentricity fluctuation, {\bf b)} matching the
multiplicity, {\bf c)} re-scaling the geometrical size $\sigma_r$,
{\bf d)} additional fluctuation not included by rotation, {\bf f)}
default Glauber geometry versus CGC geometry, {\bf g)} default
Glauber geometry versus two-component geometry.

\begin{figure}[ht]
\centering
\epsfig{file=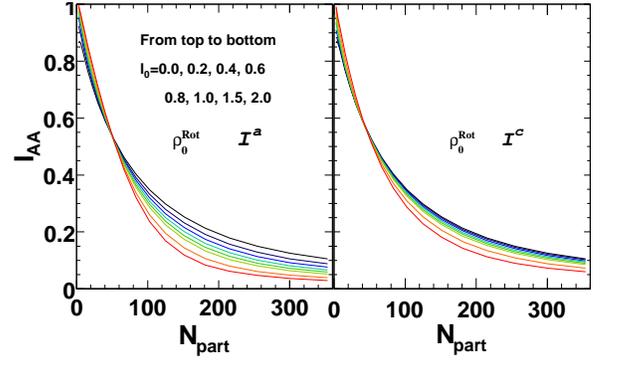,width=0.9\linewidth}
\caption{\label{fig:8d}
 (Color online) The centrality dependence of $I_{\rm AA}$ as a function of thermalization time $l_0$; Two different jet absorption schemes are used,
i.e. $I_a$ in the left panel (Eq.~\ref{eq:int1}) and $I_c$ in the right panel (Eq.~\ref{eq:int3}).
Curves that are higher and darker corresponds to smaller $l_0$ (opposite to the ordering for $v_2$, $R_{\rm AA}$ and $v_2^{I_{\rm AA}}$).}
\end{figure}

\begin{figure}[ht]
\centering
\epsfig{file=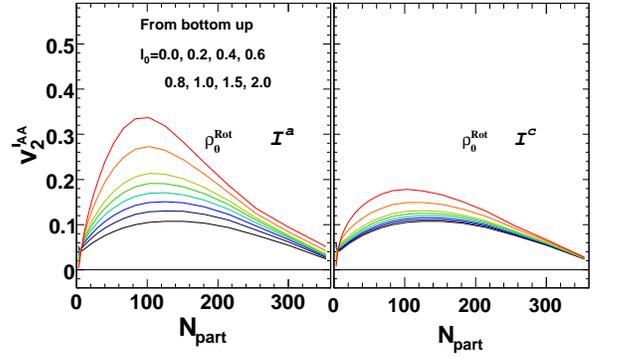,width=0.9\linewidth}
\caption{\label{fig:8e} Similar to Fig.~\ref{fig:8d}, except plotted for $v_2^{I_{\rm AA}}$.
Curves that are lower and darker corresponds to smaller $l_0$}
\end{figure}

In general, we see that $I_{\rm AA}$ and $v_2^{I_{\rm AA}}$ are
much more sensitive to their counterpart single particle
observables, $R_{\rm AA}$ and $v_2$. They are also very sensitive
to both the path-length dependence and the thermalization time
$\tau_0$ (see Figs.~\ref{fig:8d} and ~\ref{fig:8e}). In one case
(Fig.~\ref{fig:8c}c)), the $v_2^{I_{\rm AA}}$ even becomes negative
in central collisions, reflecting the dominance of tangential
emission when the collision profile is used as the matter profile
($\rho_1$). This is the case because $\rho_1$ has the narrowest
profile (15\% smaller than participant profile), such that more
jets are generated on the surface and can survive if emitted
tangentially. Unlike the $v_2$, the change in $v_2^{I_{\rm AA}}$,
when tuned to same multiplicity and average, is not proportional to
the corresponding change in $\epsilon$. The only change to which
$I_{\rm AA}$ and $v_2^{I_{\rm AA}}$ is not sensitive is the
fluctuation of the PP angle.

\section{Sensitivity to $\kappa$}
\label{app:AA} We have discussed four physics observables, $R_{\rm
AA}$, $v_2$, $I_{\rm AA}$, and $v_2^{I_{\rm AA}}$. Their centrality
dependencies and path length dependencies are entirely controlled
by the $\kappa$ parameter introduced in Sec.~\ref{sec:2}. For each
running mode, $\kappa$ is always readjusted independently such that
$R_{\rm AA}=0.182$ in the 0\%-5\% centrality bin.

The discussion so far has ignored the experimental uncertainty on
the $R_{\rm AA}$ measurement, and it is about 12\% for the 0\%-5\%
centrality bin ($0.160<R_{\rm AA}<0.204$). This uncertainty
invariability translates into some uncertainty on $\kappa$, which
in turn translates into some uncertainties on the other three
observables. Table.~\ref{tab:2} summarizes the $\kappa$ values and
associated uncertainties calculated for $\rho_0$, $\rho_2$ and
$\rho_3$ and $m=1-4$. One can see that the fractional uncertainties
increase gradually for larger $m$. The differences between
$\rho_0$, $\rho_2$ and $\rho_3$, however, can be largely attributed
to their different total matter integral in the 0\%-5\% bin, which
are 353, 301 and 695 for $\rho_0$, $\rho_2$ and $\rho_3$,
respectively. If we readjust the total matter integral for the
0\%-5\% centrality bin to the same number, say 353, then the
obtained $\kappa$ would need to be multiplied by $301/353=0.854$
for $\rho_2$ and $695/353=1.97$ for $\rho_3$, respectively. This
procedure makes the $\kappa$ rather close to each other for
different $\rho$'s; the residual differences then can be attributed
to their different shapes and sizes.

\begin{table}
\caption{\label{tab:2} The $\kappa$ values for three matter
profiles, $\rho_0$ (participant profile), $\rho_2$ (two component
profile with $\delta=0.14$), and $\rho_3$ (CGC profile), and four
$l$ dependencies ($m=1,2,3,4$). The range of $\kappa$ in each case
corresponds to $0.160<R_{\rm AA}<0.204$ in the 0\%-5\% centrality
bin. The total matter integrals for each profile are also listed.}
\small{
\begin{tabular}{l|l|l|l|l|l|l}\hline
        &Integral &$I_{1}$ &$I_{2}$ &$I_{3}$&$I_{4}$\\\hline
$\rho_0$&353 &$0.147^{+12\%}_{-10\%}$  & $0.082^{+18\%}_{-14\%}$ & $0.035^{+24\%}_{-18\%}$& $0.0125^{+29\%}_{-21\%}$\\
$\rho_2$&301 &$0.185^{+14\%}_{-11\%}$  & $0.117^{+20\%}_{-15\%}$ & $0.054^{+26\%}_{-19\%}$& $0.0206^{+31\%}_{-22\%}$\\
$\rho_3$&695 &$0.076^{+12\%}_{-10\%}$  & $0.046^{+19\%}_{-14\%}$ & $0.020^{+24\%}_{-18\%}$& $0.0073^{+29\%}_{-21\%}$\\\hline
\end{tabular}
}\normalsize
\end{table}

\begin{figure}[ht]
\centering
\epsfig{file=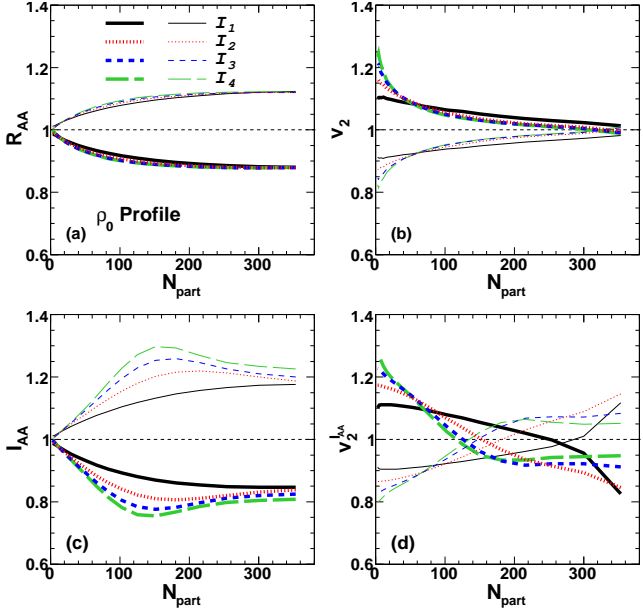,width=1.0\linewidth}
\caption{\label{fig:AA0}
 (Color online) Centrality dependence of (a): $R_{\rm AA}$, (b) $v_{2}$, (c) $I_{\rm AA}$ and (d) $v_2^{I_{\rm AA}}$
using a $\rho_0$ matter profile with $\kappa$ varied to allow the $R_{\rm
AA}$ to change by $\pm$12\% in the 0\%-5\% centrality bin. Different line styles are used for $m=1$ (solid lines),
$m=2$ (dotted lines), $m=3$ (dashed lines), and $m=4$ (long dashed lines); the upper (lower) values of $\kappa$ are represented by thin (thick) lines.
Similar results are also observed for $\rho_2$ and $\rho_3$ matter profiles (not shown).}
\end{figure}

Figure~\ref{fig:AA0} shows the centrality dependence of the four
jet quenching observables using a $\rho_0$ matter
profile~\footnote{Similar results are also observed for $\rho_2$
and $\rho_3$ matter profiles.} with $\kappa$ varied to allow the
$R_{\rm AA}$ to change by $\pm$12\% in the 0\%-5\% centrality bin
for each $m$ value (panel (a)). These $\kappa$ values are then used
to predict the centrality dependence of $v_2$ (panel (b)), $I_{\rm
AA}$ (panel (c)) and $v_2^{I_{\rm AA}}$ (panel (d)). We can see
that the fractional change on $v_2$ is typically smaller than 5\%
for $N_{\rm part}>100$, and the change decreases for larger $N_{\rm
part}$; the direction of the change, however, is anti-correlated
with the change of $R_{\rm AA}$. The fractional change on $I_{\rm
AA}$ is about 20\%, about twice amount of the change of $R_{\rm
AA}$; this change is in the same direction as that for the $R_{\rm
AA}$. The fractional change in $v_2^{I_{\rm AA}}$ is more
complicated: It typically decreases for central collisions and
increases for mid-central and peripheral collisions, but the
overall change is typically less than 5\%.

\section{Comments on finite nucleon size effect, event-by-event
fluctuation etc} \label{app:B}

\begin{figure}[ht]
\centering
\epsfig{file=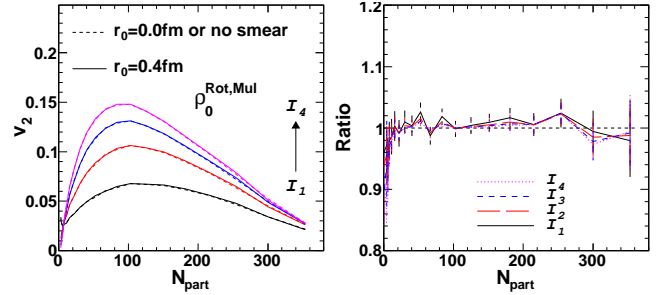,width=1.0\linewidth}
\caption{\label{fig:A1} (Color online) Comparison of the $v_2$'s for Glauber geometry, which are filled either according to center positions of nucleons
 or assuming Gauss profiles with a width of 0.4 fm.}
\end{figure}

We can show explicitly why the finite nucleon size is not important
for $v_2$ calculation except in most peripheral collisions. We
notice that the finite nucleon size leads to an increase of the
variance matter profile, but does not change the orientation of the
rotated frame:
\begin{eqnarray}
\nonumber
\sigma_x^{\prime2}&\rightarrow&\sigma_x^{\prime2}+r_0^2\\\nonumber
\sigma_y^{\prime2}&\rightarrow&\sigma_y^{\prime2}+r_0^2\\\nonumber
\sigma_{xy}^{\prime}=0&\rightarrow&\sigma_{xy}^{\prime}=0\\\nonumber
\sigma_r^{\prime2}&\rightarrow&\sigma_r^{\prime2}+r_0^2\\\nonumber
\epsilon^{\rm part}&\rightarrow&\epsilon^{\rm part}\left(1+\frac{r_0^2}{\sigma_r^2}\right)\nonumber
\end{eqnarray}
The eccentricity decreases a little bit due to smearing of $n$-$n$
overlap function. However, for a typical fireball size of
$\sigma_r=3$fm, this is only a 1.7\% change in the
eccentricity.~\footnote{The impact is somewhat larger for
$\epsilon_{\rm RP}$, i.e. average without rotation} The RMS size of
the ellipsoid also increases by a few percent, but the
corresponding jet production profile increases by the same amount,
resulting in almost no change on the calculated $v_2$.
Figure~\ref{fig:A1} shows the ratio of the $v_2$ for event averaged
$\rho_0$ calculated assuming the nucleon profile follows either the
$\delta$ function or a Gauss function with width of $r_0=0.4$ fm.
The differences of $v_2$ are well within 3\%, except at $N_{\rm
part}<20$, where the $r_0=0.4$ fm case is smaller.

\begin{figure}[ht]
\centering
\epsfig{file=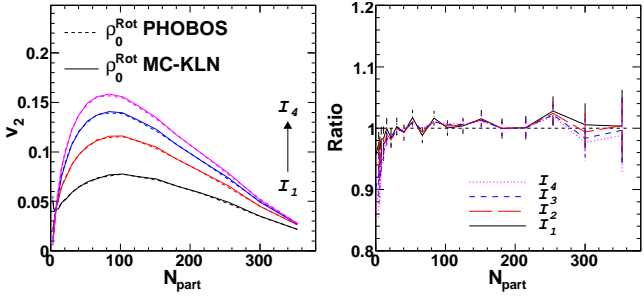,width=1.0\linewidth}
\caption{\label{fig:A0} (Color online) Comparison of the $v_2$'s for Glauber geometry calculated using the PHOBOS code and the MC-KLN code.}
\end{figure}

Note that similar findings also appear in MC-KLN, which assumes
that participating nucleons are disks of finite size in filling the
participant profile distribution (in addition to assuming finite
nucleon size for determining whether it is a participant). This is
illustrated by Fig.~\ref{fig:A0}, which compares the $v_2$ results
for the same Glauber geometry determined with either the PHOBOS
code or the MC-KLN code. The ratio drops at $N_{\rm part}<20$
similar to Fig.~\ref{fig:A1}. As a side note, we point out that
this plot also confirmed the consistency between the PHOBOS code
and the MC-KLN code for calculating the Glauber geometry when
running with the same parameters.

\begin{figure}[ht]
\centering
\epsfig{file=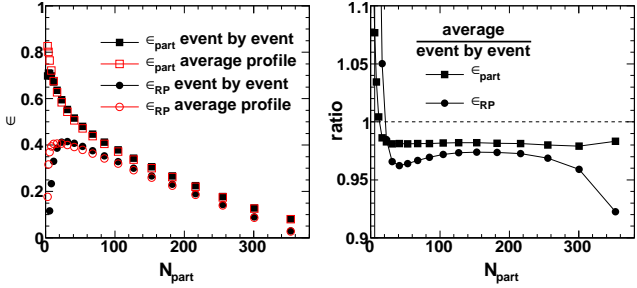,width=1\linewidth}
\caption{\label{fig:A3} (Color online) Left panel: the standard (circles) and participant (boxes) eccentricity either by averaging over their event-by-event value (filled symbols)
or calculated directly from the averaged matter profiles (open symbols). Right panel: the corresponding ratios for the standard and participant eccentricities.}
\end{figure}

\begin{figure}[ht]
\centering \epsfig{file=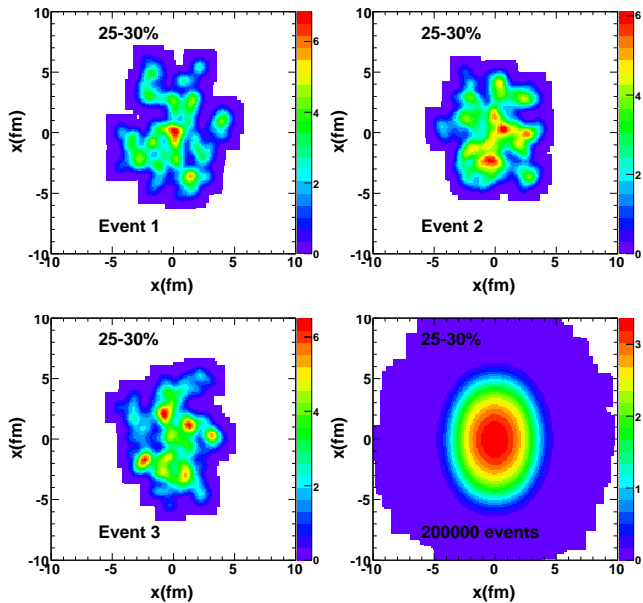,width=1\linewidth}
\caption{\label{fig:A2} (Color online) The event-by-event participant profiles ($\rho_0^{\rm Rot}$) for three typical events and profile averaged over 200,000 events for 25\%-30\% centrality selection.
Nucleons are assumed to have a Gaussian profile with a width of $r_0 = 0.4$ fm. All events have been shifted to center of gravity and
rotated to participant plane.}
\end{figure}

We have seen that the event-by-event fluctuation leads to large
dispersion of the distribution of various geometrical variables,
such as eccentricity and RMS size in Fig.~\ref{fig:5e}. Yet the
mean values seem to be insensitive to whether they are calculated
for each event and then averaged over many events or are calculated
directly from the averaged profile. As an example,
Fig.~\ref{fig:A3} compares the eccentricity averaged over values
calculated event by event $\left<\frac{\sigma_y^{\prime
2}-\sigma_x^{\prime 2}}{\sigma_y^{\prime 2}+\sigma_x^{\prime
2}}\right>$ with the eccentricity calculated from the average
profile $\frac{\langle\sigma_y^{\prime
2}\rangle-\langle\sigma_x^{\prime
2}\rangle}{\langle\sigma_y^{\prime
2}\rangle+\langle\sigma_x^{\prime 2}\rangle}$. The ratios are shown
in the right panel for both $\epsilon_{\rm RP}$ and $\epsilon_{\rm
part}$. As one can see, the difference is $<2\%$ for $\epsilon_{\rm
part}$ and somewhat larger for $\epsilon_{\rm RP}$~\footnote{ The
agreement generally worsens when the width of the distribution
becomes large relative to the mean value, which is the case for
$\epsilon_{\rm RP}$.}. This is quite remarkable given the sharp
visual contrast between the lumpiness of the event-by-event profile
and the smoothness of the average profile (see Fig.~\ref{fig:A2}).

Nevertheless, the good news is that the average profile seems to
preserve most of the relevant geometrical information. For example,
one can calculate the eccentricity directly from the overall matter
profile instead of calculating it event by event and then averaging
it over many events.

\end{document}